\documentclass[draftcls, onecolumn]{./IEEEtran}

\ifCLASSINFOpdf
  \usepackage[pdftex]{graphicx}
\else
\fi
\usepackage{amsmath}
\usepackage{amssymb}
\usepackage{amsfonts}
\usepackage{subfig}
\usepackage[colorlinks=true,
  pdfstartview=FitV,
  unicode,
  pdftitle={hierarchical modulation},
  pdfauthor={Hugo Meric and Jerome Lacan},
  pdfkeywords={hierarchical modulation}]{hyperref}

\begin{document}
%
\title{Generic Approach for Hierarchical Modulation Performance Analysis: Application to DVB-SH and DVB-S2}

%
%
%

        

\author{\IEEEauthorblockN{Hugo M{\'e}ric\IEEEauthorrefmark{1}\IEEEauthorrefmark{2},
J{\'e}r{\^o}me Lacan\IEEEauthorrefmark{2}\IEEEauthorrefmark{1},
Caroline Amiot-Bazile\IEEEauthorrefmark{3}, 
Fabrice Arnal\IEEEauthorrefmark{4} and
Marie-Laure Boucheret\IEEEauthorrefmark{2}\IEEEauthorrefmark{1}
\IEEEcompsocitemizethanks{The material in this paper will be presented in part at WTS 2011, New-York, United States, April 2011. It is available online at http://arxiv.org/abs/1103.1305.}}

\IEEEauthorblockA{\IEEEauthorrefmark{1}T{\'e}SA, Toulouse, France\\}
\IEEEauthorblockA{\IEEEauthorrefmark{2}Universit{\'e} de Toulouse, Toulouse, France\\}
\IEEEauthorblockA{\IEEEauthorrefmark{3}CNES, Toulouse, France\\}
\IEEEauthorblockA{\IEEEauthorrefmark{4}Thales Alenia Space, Toulouse, France\\
Email: hugo.meric@isae.fr, jerome.lacan@isae.fr, caroline.amiot-bazile@cnes.fr, fabrice.arnal@thalesaleniaspace.com, marie-laure.boucheret@enseeiht.fr}}

\maketitle

\begin{abstract}
Broadcasting systems have to deal with channel variability in order to offer the best rate to the users. Hierarchical modulation is a practical solution to provide different rates to the receivers in function of the channel quality. Unfortunately, the performance evaluation of such modulations requires time consuming simulations. We propose in this paper a novel approach based on the channel capacity to avoid these simulations. The method allows to study the performance of hierarchical and also classical modulations combined with error correcting codes. We will also compare hierarchical modulation with time sharing strategy in terms of achievable rates and indisponibility. Our work will be applied to the DVB-SH and DVB-S2 standards, which both consider hierarchical modulation as an optional feature.
\end{abstract}



%
\IEEEpeerreviewmaketitle

\section{Introduction}

In most broadcast applications, all the receivers do not experience the same signal-to-noise ratio (SNR). For instance, in satellite communications  the channel quality decreases with the presence of clouds in Ku or Ka band, or with shadowing effects of the environment in lower bands.

The first solution for broadcasting is to design the system for the worst-case reception. However, this solution does not take into account the variability of channel qualities. This holds a loss of spectrum efficiency for users with good reception. Then, two other schemes have been proposed in \cite{cover} and \cite{bergmans} to improve the first one: time division multiplexing with variable coding and modulation, and superposition coding. Time division multiplexing consists in using a first couple modulation/coding rate during a fraction of time, and then using an other modulation/coding rate for the remaining time. All the population can receive the first part of the signal called the High Priority (HP) signal and only the receivers in good conditions receive the second one called the Low Priority (LP) signal.

\begin{figure}[!ht]
\centering
\includegraphics[width = 0.5\columnwidth]{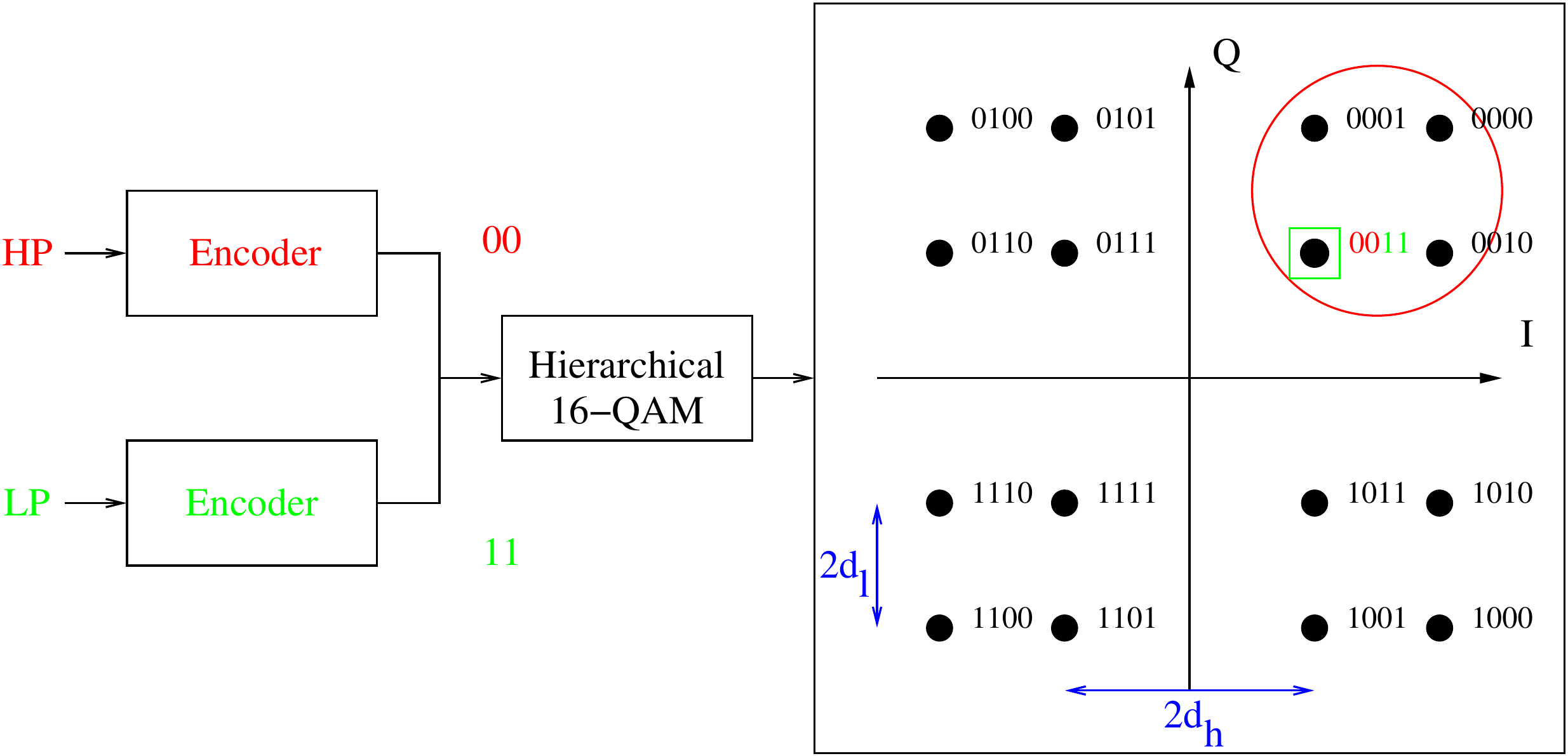}
\caption{Hierarchical Modulation using a non-uniform 16-QAM}
\label{hm_principle}
\end{figure}

Unlike time sharing, superposition coding sends information for all the receivers all the time. This scheme was shown to be optimal for the continuous Gaussian channel \cite{bergmans}. In superposition coding, the available energy is shared to several service flows which are send simultaneously and in the same band. Hierarchical modulation is a practical implementation of superposition coding. Figure~\ref{hm_principle} presents the principle of the hierarchical modulation with a non-uniform 16-QAM. The idea is to merge two different streams at the modulation step. The HP stream is used to select the quadrant, and the LP stream selects the position inside the quadrant. In good conditions receivers can decode both streams, unlike bad receivers which only locate the quadrant and then decode the HP stream as a QPSK constellation. At the service level, the two streams are either \emph{dependent} or \emph{independent}. A practical example of dependent streams is H.264/SVC encoded video. This standard generates several video layers, where each layer improves the video quality but requires all the underlying layers. Scalable Video Coding can be used with time division multiplexing \cite{svc_vcm} or hierarchical modulation \cite{svc_hm}. In this paper, we suppose that the streams are \emph{dependent} although the study can be extended to the other case.

Even if hierarchical modulation is not a new concept, its use in most recent broadcast satellite standard, DVB-SH \cite{sh} and DVB-S2 \cite{s2}, has motivated its analysis and development by the satellite community research. This article is focused on the performance analysis of hierarchical modulation and the comparison with time sharing. We propose here a new approach to evaluate the performance of hierarchical modulation. The method is based on the channel capacity and relies on the fact that the real code at coding rate $R$ is similar to a theoretical ideal code at coding rate $\tilde R$ in terms of decoding threshold. Our approach is applied to DVB-SH and DVB-S2, where it helps to decide the good coding strategy for a system with constraints.

The paper presents our work as follows:
\begin{itemize}
\item In Section~\ref{part2}, we compute the capacity for any hierarchical modulation. A first comparison between time sharing and hierarchical modulation is done by comparing their set of theoritical achievable rates, called the capacity region.
\item In Section~\ref{part3}, we propose a method using the capacity to evaluate the performance of hierarchical modulations in terms of spectral efficiency and required $E_s/N_0$ for a targeted Bit Error Rate (BER) or Packet Error Rate (PER). Then, we obtain the achievable rates for real codes using hierarchical modulation or time sharing.
\item Section~\ref{part4} introduces the notion of indisponibility which requires SNR distribution. This allows to complete our comparison between the two studied schemes.
\end{itemize}
Section~\ref{conclusion} concludes the paper by summarizing the results.

\section{Capacity of the Hierarchical Modulation}\label{part2}

This section defines and computes the channel capacity for any hierarchical modulation. Then, the capacity is used to compare two different schemes: time division multiplexing and superposition coding. Our performance analysis method presented in Section~\ref{part3} is also based on the capacity.

\subsection{Computation of the capacity}
A channel can be considered as a system consisting of an input alphabet, an output alphabet and a probability transition matrix $p(y|x)$. We define two random variables $X$, $Y$ representing the input and output alphabets respectively. The mutual information between $X$ and $Y$, noted $I(X;Y)$, measures the amount of information conveyed by Y about X. For two discrete random variables $X$ and $Y$, the expression of the mutual information is
\begin{eqnarray}
I(X;Y) = \sum_{x \in \mathcal{X}} \sum_{y \in \mathcal{Y}} p(x) p(y|x) \log_{2} \left( \frac{p(y|x)}{p(y)} \right).
\label{defmi}
\end{eqnarray}

Using this notion, the channel capacity is then given by
\begin{eqnarray}
C = \max_{p(x)} I(X;Y) ,
\label{mi}
\end{eqnarray}
where the maximum is computed over all possible input distributions \cite{cove_thom_91}.

Here we consider the memoryless discrete input and continuous output Gaussian channel. The discrete inputs $\mathbf{x}_i$ are obtained using a modulation and belong to a set of discrete points $\chi \subset \mathbb{R}^2$ of size $|\chi|=M=2^m$ called the constellation. Thus, each symbol of the constellation carries m bits. Reference \cite[Chapter 3]{berrou} gives an explicit formula (\ref{capacity_berrou}) for the capacity in this particular case.
\begin{eqnarray}
C = \log_{2}(M) - \frac{1}{M} \sum_{i=1}^M \int\limits_{-\infty}^{+\infty} \int\limits_{-\infty}^{+\infty} p(\mathbf{y}|\mathbf{x}_i) \log_{2} \left( \frac{\sum_{j=1}^M p(\mathbf{y}|\mathbf{x}_j)}{p(\mathbf{y}|\mathbf{x}_i)}\right) \, \mathrm d\mathbf{y}
\label{capacity_berrou}
\end{eqnarray}

Equation~(\ref{capacity_berrou}) computes the capacity for one stream using all the bits. We are now interested to evaluate the capacity of a stream using a \emph{subset} of the m bits. Reference \cite{caire} presents the case where each stream uses one bit. The idea is to modify the input random variable $X$ by the input bits used in each stream. We define $b_i$ as the value of the i\textsuperscript{th} bit of the label of any constellation point $\mathbf{x}$. Suppose the stream uses \emph{k bits among m} in the positions $j_1$, ..., $j_k$. For any integer i, let $l_n(i)$ denote the n\textsuperscript{th} bit in the binary representation of i such as $i=\sum_{n=1}^{+\infty} l_n(i)2^{n-1}$. Then using (\ref{defmi}) with the new input variable and continuous output, the capacity of the stream is
\begin{eqnarray}
C = \frac{1}{2^k} \sum_{i=0}^{2^k-1} \int\limits_{-\infty}^{+\infty} \int\limits_{-\infty}^{+\infty} \underbrace{p\left(\mathbf{y}|b_{j_1}=l_1(i),...,b_{j_k}=l_k(i) \right)}_{\text{sum over all possible k-uplets}}  \log_{2} \left( \frac{ p\left(\mathbf{y}|b_{j_1}=l_1(i),...,b_{j_k}=l_k(i) \right) }{ p(\mathbf{y}) }\right) \, \mathrm d\mathbf{y} ,
\label{capacity_hm}
\end{eqnarray}
where $p(\mathbf{y}) = \frac{1}{2^k} \sum_{i=0}^{2^k-1} p\left(\mathbf{y}|b_{j_1}=l_1(i),...,b_{j_k}=l_k(i) \right)$.

Let $L_n(\mathbf{x})$ denote the n\textsuperscript{th} bit of the label of any constellation point $\mathbf{x}$. We introduce $\chi_i$ the subset of $\chi$ defined as follows
\begin{eqnarray}
 \chi_i = \{ \mathbf{x} \in \chi | L_{j_1}(\mathbf{x})=l_1(i) , ... , L_{j_k}(\mathbf{x})=l_k(i) \}.
\label{chi}
\end{eqnarray}
 
\begin{figure}[!ht]
\centering
\includegraphics[width = 0.35\textwidth]{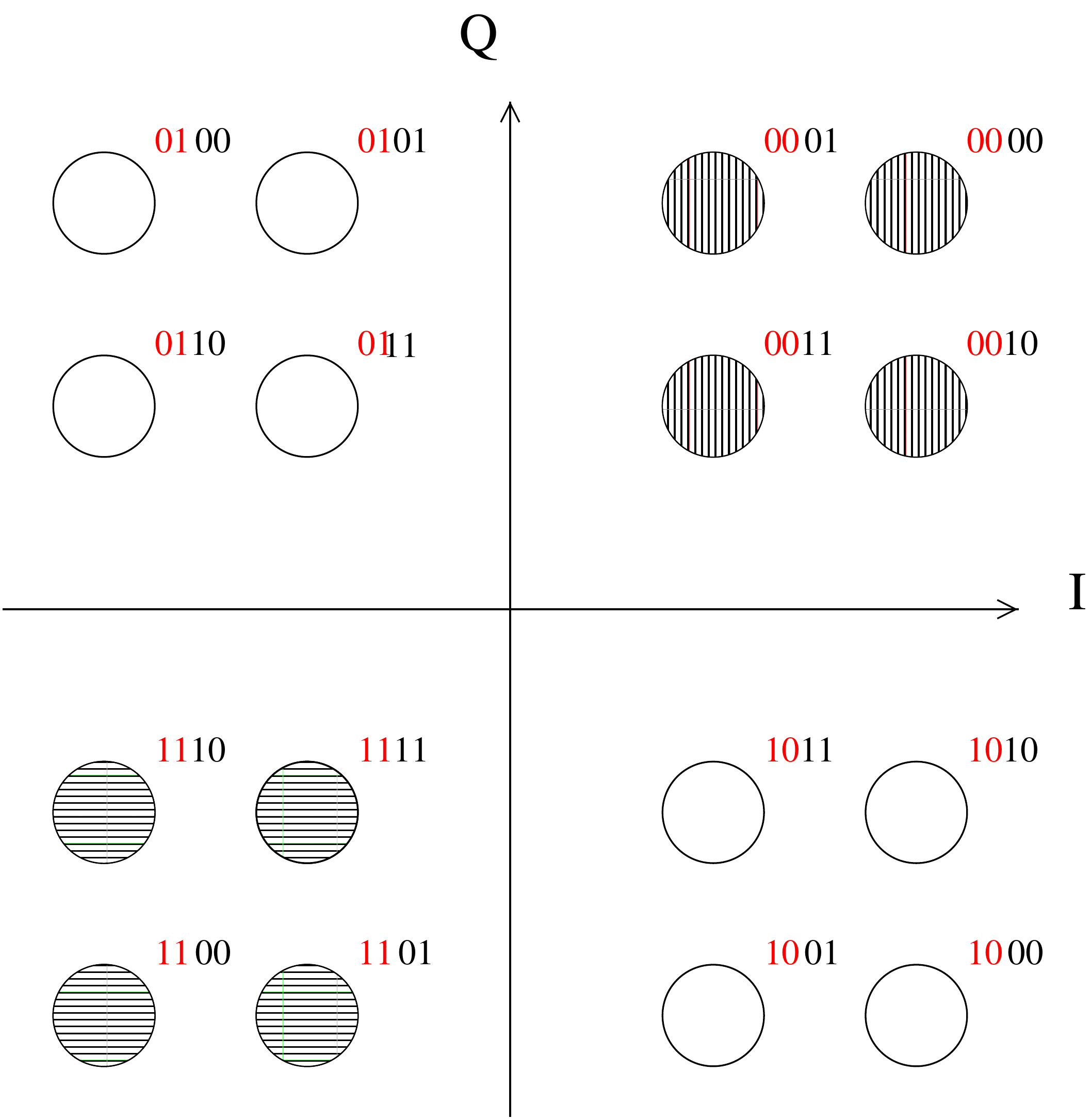}
\caption{Examples of $\chi_0$ (vertical lines) and $\chi_3$ (horizontal lines)}
\label{subset}
\end{figure}

The set $\chi_i$ depends on i and the positions of the bits involved in the stream. Figure~\ref{subset} shows an example of subsets for a 16-QAM with a particular mapping, where the stream uses bits 1 and 2 (in that case $k=2$, $j_1=1$ and $j_2=2$).  

Then the conditional probability density function of $\mathbf{y}$ in (\ref{capacity_hm}) can be written
\begin{eqnarray}
p\left(\mathbf{y}|b_{j_1}=l_1(i),...,b_{j_k}=l_k(i) \right) & = & 
\sum_{\mathbf{x} \in \chi_i} p(\mathbf{y}|\mathbf{x}) p(\mathbf{x}|\mathbf{x} \in \chi_i) \nonumber\\
&=& \frac{1}{|\chi_i|} \sum_{\mathbf{x} \in \chi_i} p(\mathbf{y}|\mathbf{x}) ,
\label{conditional_prob}
\end{eqnarray}

where $|\chi_i|=2^{m-k}$ for all i. Moreover, the transition distribution $p(\mathbf{y}|\mathbf{x})$ for a Gaussian channel is 
\begin{eqnarray}
p(\mathbf{y}|\mathbf{x}) = \frac{1}{\pi N_0} \exp \left( -\frac{\| \mathbf{y}-\mathbf{x} \|^2}{N_0} \right).
\label{transition_distribution}
\end{eqnarray}

Using (\ref{conditional_prob}) and (\ref{transition_distribution}) in (\ref{capacity_hm}), we finally obtain the capacity for one stream (\ref{final_capacity_hm}). The capacity is an increasing function of $E_s/N_0$ and its value is less or equal to k the number of bits used in the stream. The positivity of the capacity results of its mathematical definition \cite{cove_thom_91}.
\setlength{\arraycolsep}{0.0em}
\begin{eqnarray}
C = k &{}-{}& \frac{1}{2^k\pi}\sum_{i=0}^{2^k-1} \int\limits_{-\infty}^{+\infty} \int\limits_{-\infty}^{+\infty} \left( \frac{1}{|\chi_i|}\sum_{\mathbf{x} \in \chi_i} \exp \left( \begin{Vmatrix} \mathbf{u}-\frac{\mathbf{x}}{\sqrt{N_0}} \end{Vmatrix}^2 \right) \right) \log_{2} \left( 1  + \frac{\displaystyle \sum_{\mathbf{x} \in \chi \setminus \chi_i} \exp \left( \begin{Vmatrix} \mathbf{u}-\frac{\mathbf{x}}{\sqrt{N_0}} \end{Vmatrix}^2 \right)}{\displaystyle \sum_{\mathbf{x} \in \chi_i} \exp \left( \begin{Vmatrix} \mathbf{u}-\frac{\mathbf{x}}{\sqrt{N_0}} \end{Vmatrix}^2 \right)}\right) \, \mathrm d\mathbf{u} 
\label{final_capacity_hm}
\end{eqnarray}
\setlength{\arraycolsep}{5pt}

\subsection{Case of non-uniform hierarchical modulation capacities}

We begin with few definitions before applying (\ref{final_capacity_hm}) to the hierarchical 16-QAM and 8-PSK considered in DVB-SH and DVB-S2 respectively.

\begin{figure}[!ht]
\centering
\includegraphics[width = 0.25\columnwidth]{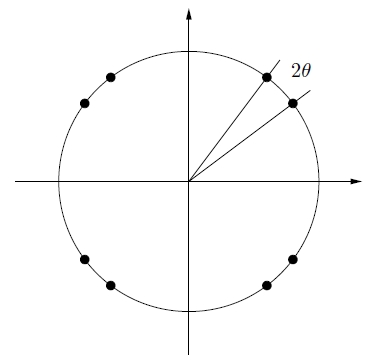}
\caption{Non-uniform 8-PSK}
\label{nupsk}
\end{figure}

Hierarchical modulations merge several streams in a same symbol. They often use non-uniform constellation. Non-uniform constellations are opposed to uniform constellations, where the symbols are uniformly distributed. The constellation parameter is defined to describe non-uniform constellations. Figure~\ref{hm_principle} illustrates a non-uniform 16-QAM. The constellation parameter $\alpha$ is defined by $\alpha=d_h/d_l$, where $2d_h$ is the minimum distance between two constellation points carrying different HP bits, and $2d_l$ is the minimum distance between any constellation point. Typically, we have $\alpha \ge 1$, where $\alpha=1$ corresponds to the uniform 16-QAM, but it is also possible to have $\alpha \le 1$. DVB-SH standard recommends two values for $\alpha$: 2 and 4. DVB-S2 considers the non-uniform 8-PSK presented on Figure~\ref{nupsk}. The constellation parameter $\theta$ represents the half angle between two points in one quadrant and is selected by the operator according to the desired performance. In both cases, the constellation parameter has a great impact on the performance of the decoded stream.

\begin{figure*}[!ht]
\centerline{\subfloat[$\alpha=2$]{\includegraphics[width=0.5\columnwidth]{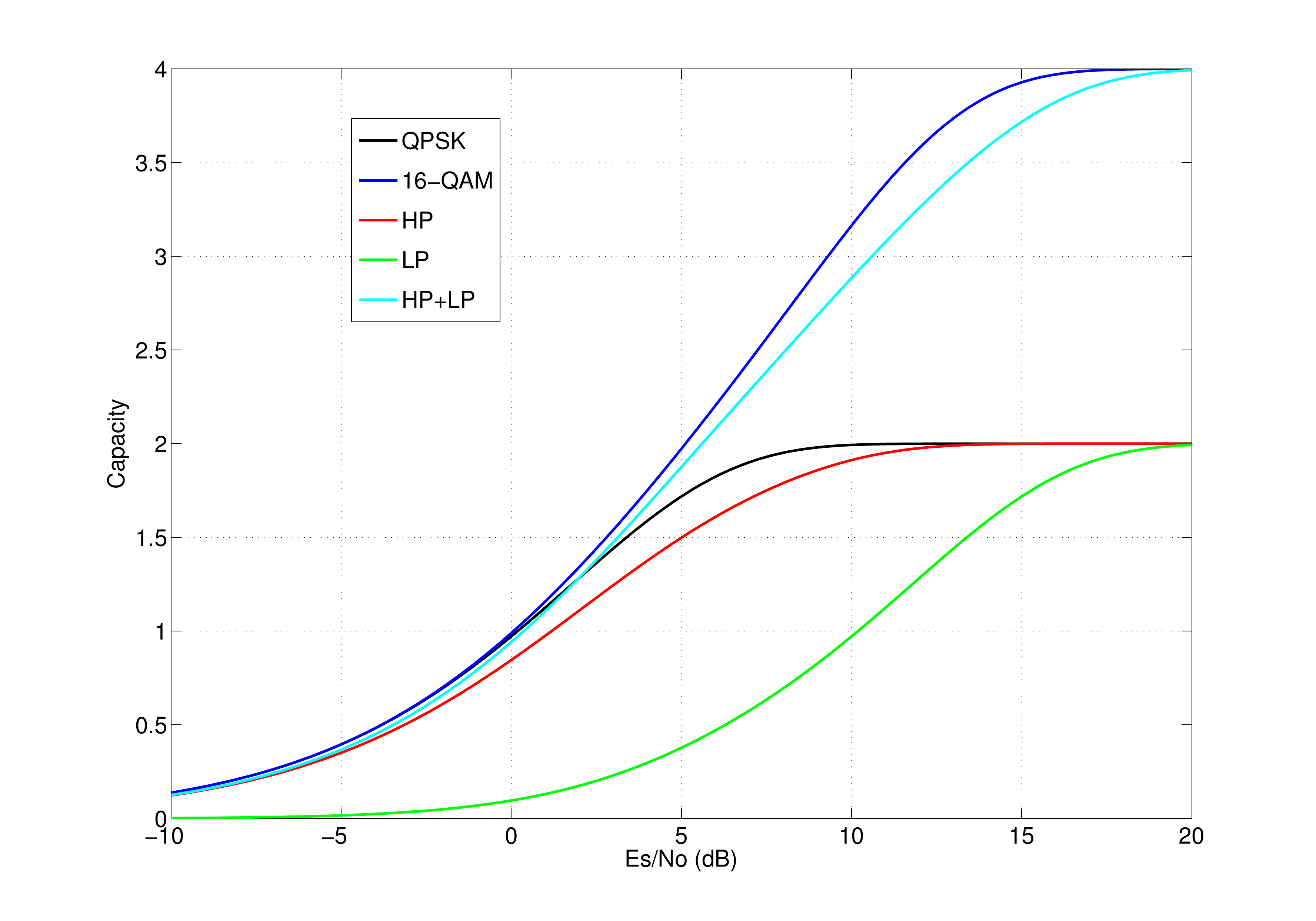}%
\label{capacity_2}}%
\hfil
\subfloat[$\alpha=4$]{\includegraphics[width=0.5\columnwidth]{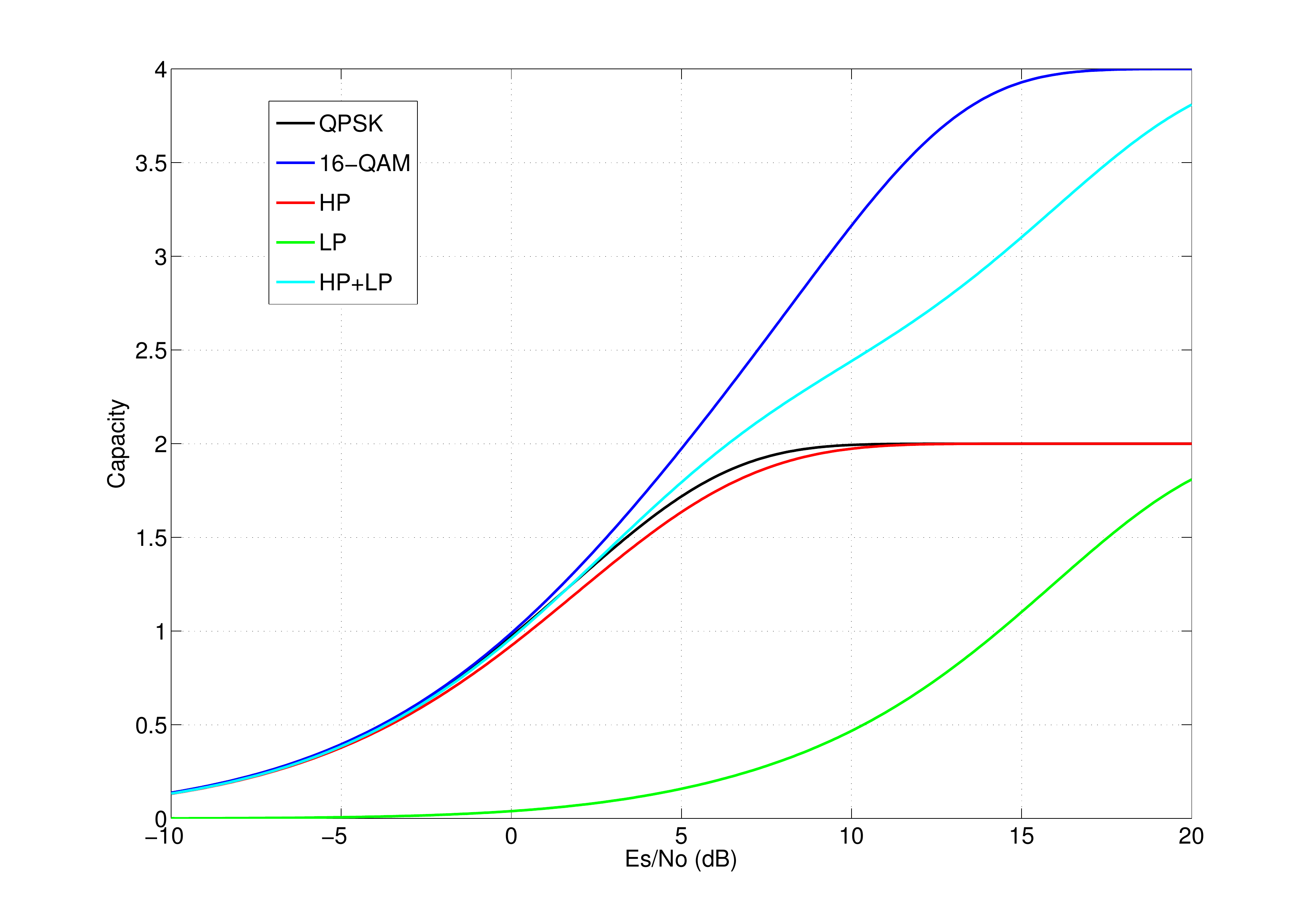}%
\label{capacity_4}}}%
\caption{16-QAM Hierarchical Modulation Capacity}
\label{capacity_sh}
\end{figure*}

\begin{figure*}[!ht]
\centerline{\subfloat[$\theta=10^\circ$]{\includegraphics[width=0.5\columnwidth]{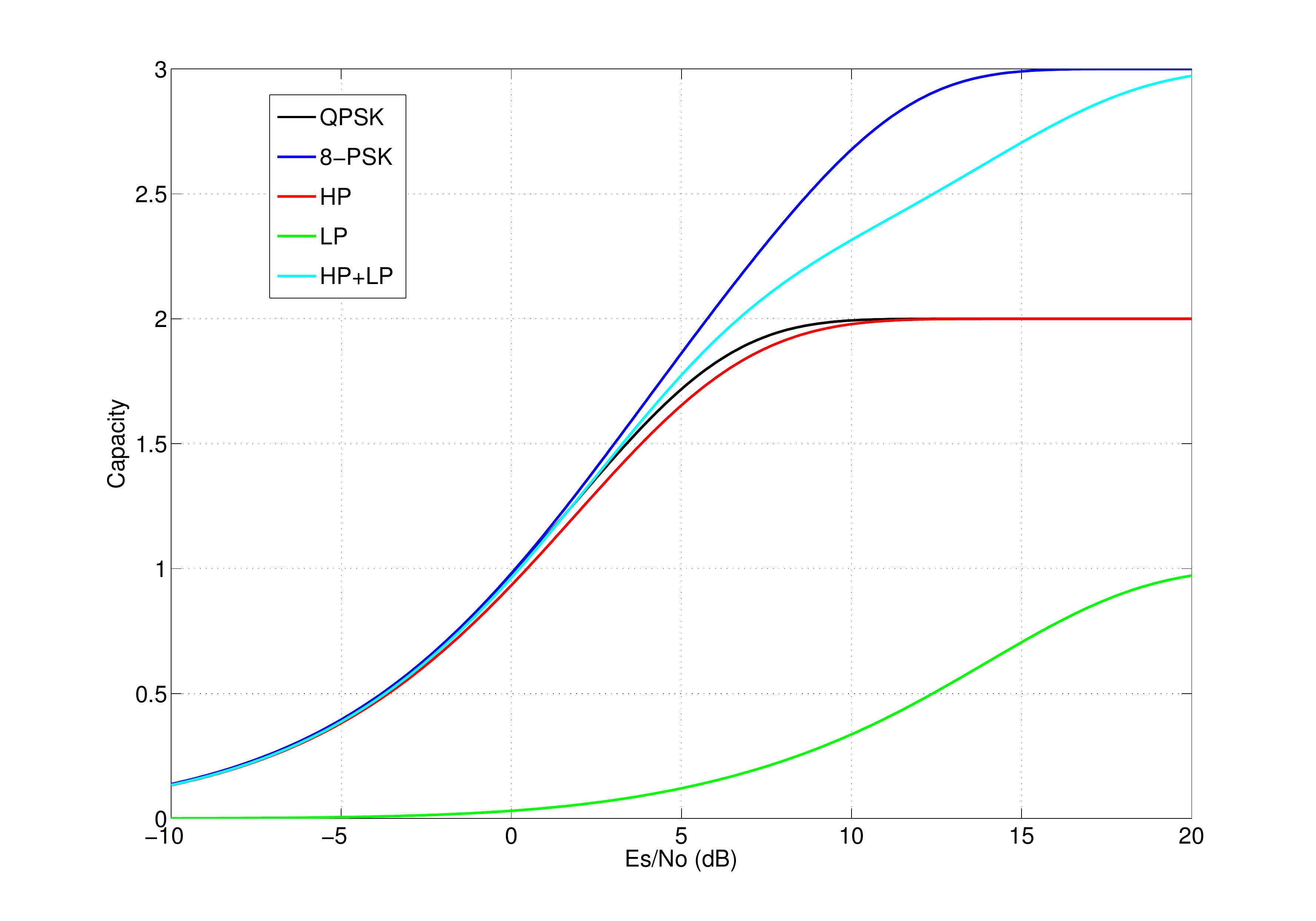}%
\label{capacity_10}}%
\hfil
\subfloat[$\theta=15^\circ$]{\includegraphics[width=0.5\columnwidth]{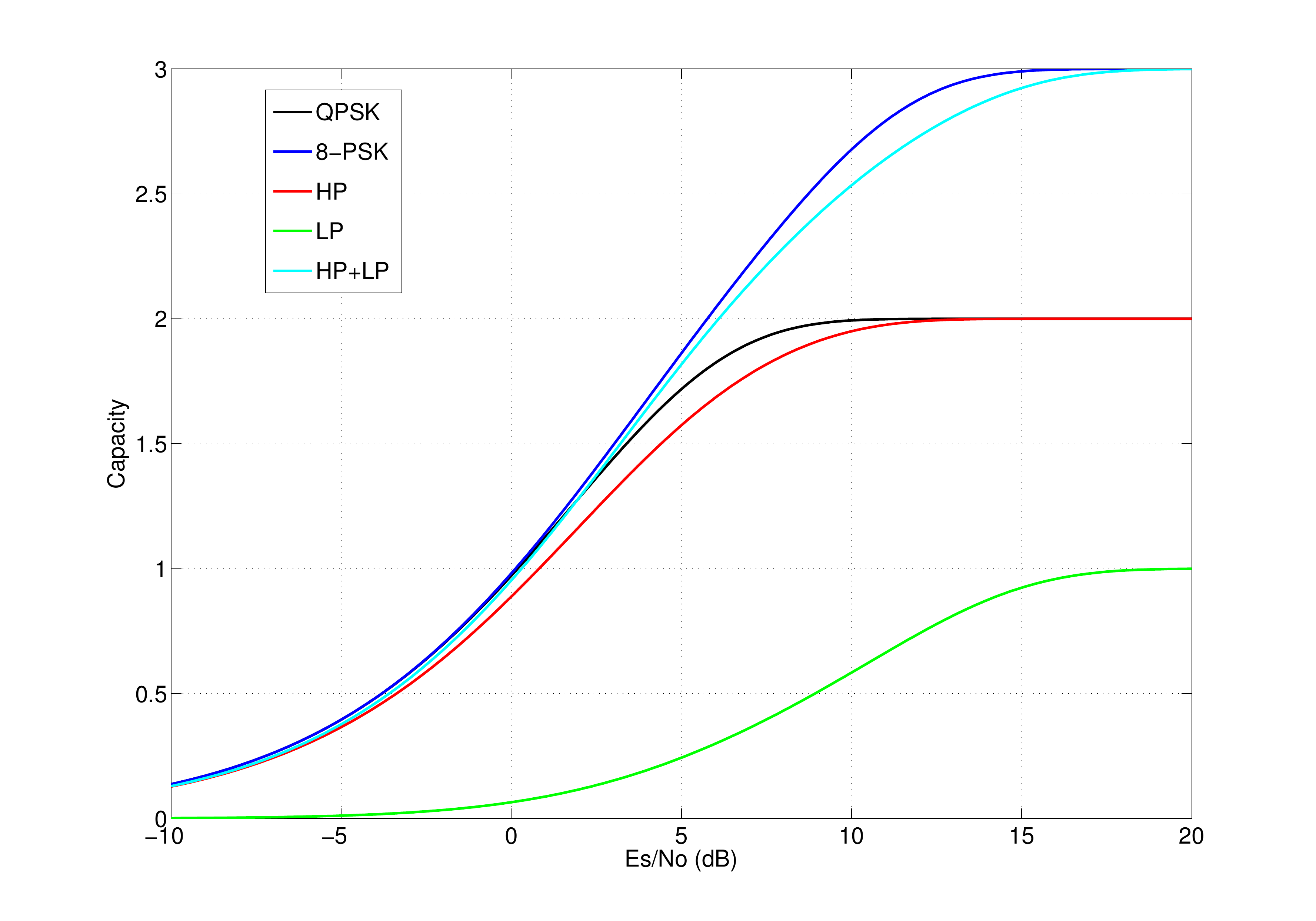}%
\label{capacity_15}}}%
\caption{8-PSK Hierarchical Modulation Capacity}
\label{capacity_s2}
\end{figure*}

Back to the hierarchical modulation capacity, we suppose the HP stream uses the bits in position 1 and 2 in both cases. The LP stream involves the remaining bits, i.e., bits 3 and 4 for the 16-QAM and bit 3 for the 8-PSK. From these definitions, we can apply (\ref{final_capacity_hm}) to compute their capacity. Figure~\ref{capacity_sh} presents the capacity for the 16-QAM for the two values of $\alpha$ defined in the DVB-SH guidelines \cite{sh} and Figure~\ref{capacity_s2} shows the results for the 8-PSK.

In Figure~\ref{capacity_sh}, when $\alpha$ grows, the constellation points in one quadrant become closer and then it is natural than the capacity for the LP stream decreases at a given SNR as it is harder to decode. For the HP stream the points are farer to the I and Q axes, then it is easier to decode which quadrant was sent and the capacity is increased. This capacity is nevertheless limited by the QPSK capacity. The same remarks are valid concerning $\theta$ in Figure~\ref{capacity_s2}. When $\theta$ decreases, it is easier to decode the good quadrant but not the symbol sent. In these examples, we apply (\ref{final_capacity_hm}) to hierarchical modulation with two flows, but it can also be applied to multilevel hierarchical modulation such as 256-QAM with 3 flows, where each flow is composed of two bits.

To conclude, we can observe that the hierarchical modulation capacity (i.e., HP+LP) is always less than the 16-QAM one for any value of $\alpha > 1$. This result has been proved in \cite{caire}, where the different streams use one bit. We can also note that the best total capacity is achieved when the constellation is close to the uniform constellation, but on the other hand performances for the HP stream are decreased. 

\subsection{Time Sharing vs Hierarchical Modulation}

We consider in this section the problem of broadcasting data to two populations, each one with a target SNR. We suppose to transmit \emph{dependent} data, for instance two layers of a video encoded with H.264/SVC. Classicaly, broadcast channels can use time sharing and/or hierarchical modulation. The Gaussian channel was studied in \cite{cover} and  \cite{bergmans}, where the superposition coding was introduced and shown to be optimal in terms of achievable rates. We investigate hereafter the case of the memoryless discrete input and continuous output Gaussian channel by computing the capacity region for the hierarchical modulation and the time sharing strategies.

Variable Coding Modulation (VCM) is a practical implementation of time sharing. During a fraction of time, a first modulation and coding rate is selected, then for the remaining time an other modulation and coding scheme is used. When VCM uses two QPSK, the corresponding modulation for superposition coding is a hierarchical 16-QAM. We are interested to compare the achievable rates between these two schemes: VCM using two QPSK  and hierarchical 16-QAM. The power allocation to each stream determines the value of $\alpha$. We suppose to have a total power budget E. The power allocation for the HP stream is given by the energy of a QPSK with a parameter $d_h+d_l$: $E_{hp}=2(d_h+d_l)^2=2d_l^2(1+\alpha)^2=\rho_{hp} E$. For the LP stream, the remaining energy is $E_{lp}=2d_l^2=\rho_{lp} E$ (QPSK with parameter $d_l$), where $\rho_{lp} = 1-\rho_{hp}$. The relation between the power allocation and $\alpha$ is
\begin{eqnarray}
\frac{E_{hp}}{E_{lp}} = \frac{\rho_{hp}}{\rho_{lp}} = (1+\alpha)^2 .
\label{alpha}
\end{eqnarray}

\begin{figure*}[!ht]
\centerline{\subfloat[SNR$_1$=2dB, SNR$_2$=10dB]{\includegraphics[width=0.5\columnwidth]{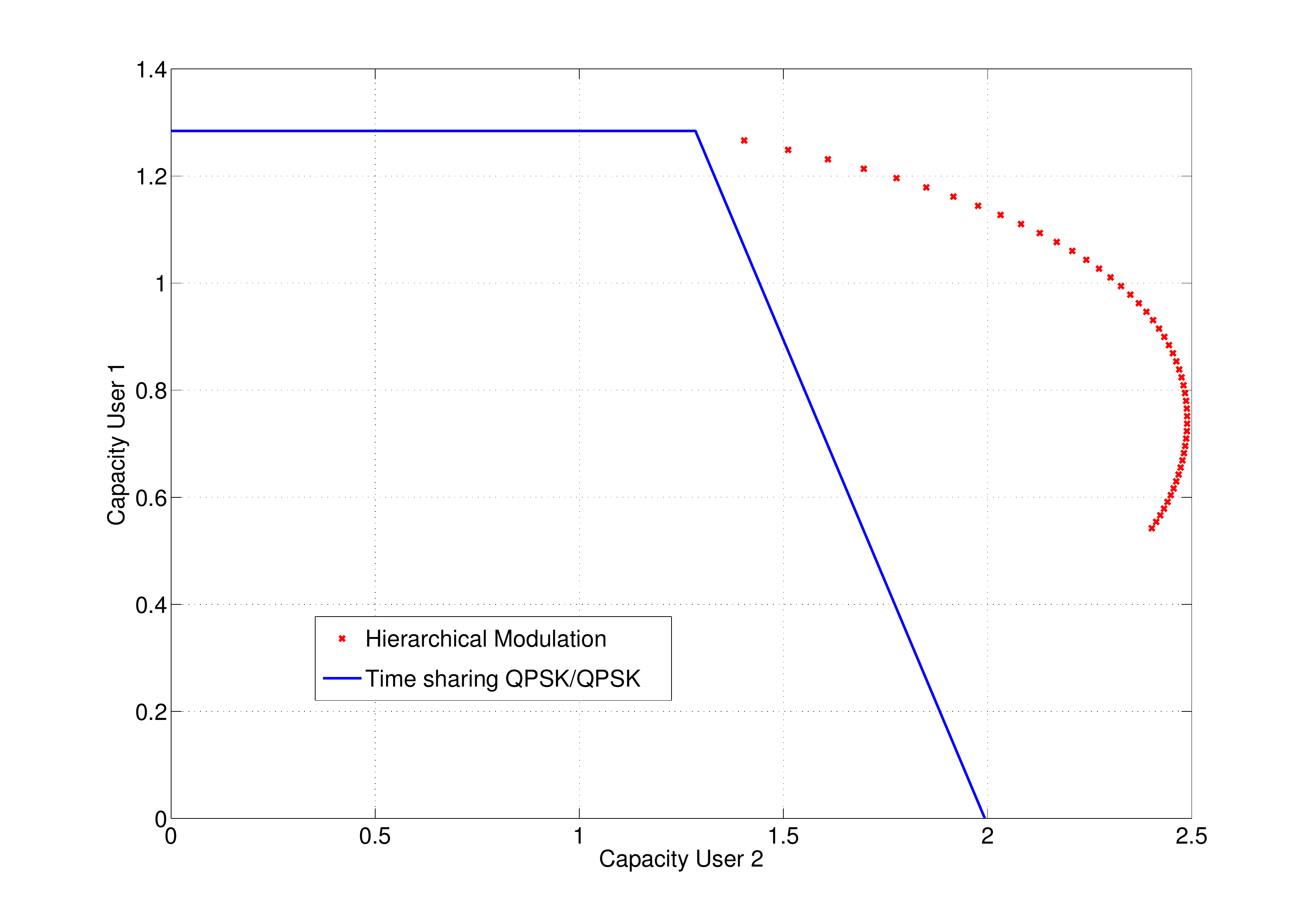}%
\label{superposition_2_10}}%
\hfil
\subfloat[SNR$_1$=-2dB, SNR$_2$=6dB]{\includegraphics[width=0.5\columnwidth]{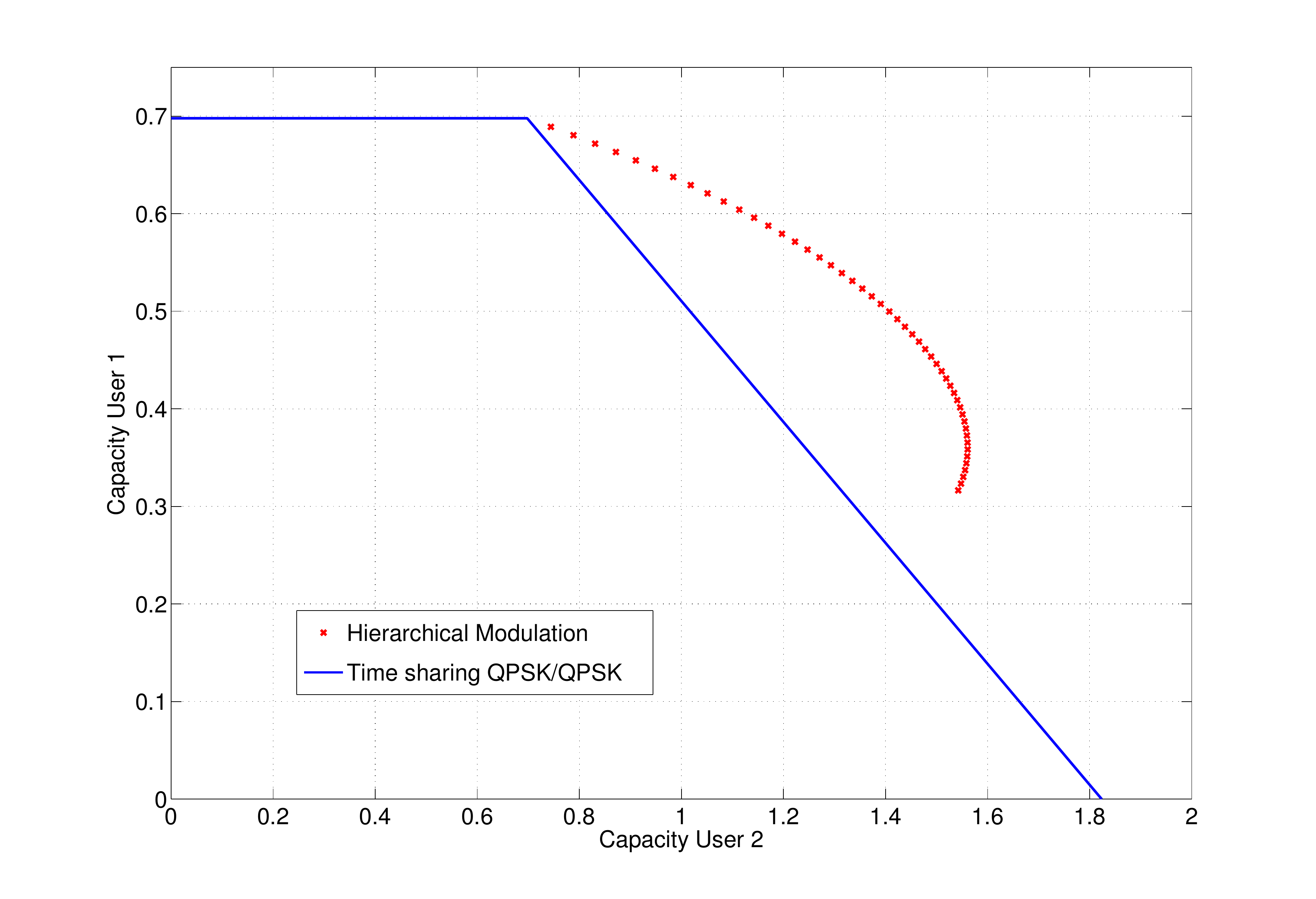}%
\label{superposition_-2_6}}}%
\caption{Capacity Region}
\label{capacity_region}
\end{figure*}

Figure~\ref{capacity_region} presents the capacity regions for two populations of users with two SNR configurations. $SNR_i$ corresponds to the SNR experienced by the users of the population i, where i equals 1 or 2. In both cases, the SNR is better for users of population 2. As mentioned previously, we suppose that the content sent to both populations is \emph{dependent}, then the achievable rates are,
\begin{eqnarray}
 R_1 &=& C_{hp} \left( SNR_1 \right) , \nonumber\\
 R_2 &=& C_{hp}\left( SNR_1 \right) + C_{lp}\left( SNR_2 \right).
\label{achievable_rates}
\end{eqnarray}

Each point on Figure~\ref{capacity_region} corresponds to a specific power configuration and then a particular value of $\alpha$ as shown in (\ref{alpha}). For the simulations, $\rho_{hp}$ varies from 0.51 to 0.99, which corresponds to a variation of $\alpha$ from 0.02 to 8.95. We now consider the curve concerning the hierarchical modulation on the Figure~\ref{superposition_2_10}. The more energy is allocated to the HP stream, higher is the capacity for population 1. When we decrease the HP stream energy, it also decreases the capacity of population 1, but the capacity of population 2 can be increased. On both figures, the hierarchical modulation outperforms the time sharing strategy. However, the achievable rate for the population 2 using hierarchical modulation can not always reach the maximum rate given by the time sharing strategy as on Figure~\ref{superposition_-2_6}.

Finally, we have seen on Figure~\ref{capacity_region} the capacity region for the hierarchical modulation. Moreover, when two sets of rates  $(R_1,R_2)$ and $\left(\tilde{R_1},\tilde{R_2}\right)$ are achievable, the time sharing strategy allows any rate pair 
\begin{eqnarray}
\left(\tau R_1+(1-\tau)\tilde{R_1} , \tau R_2+(1-\tau)\tilde{R_2} \right), 0 \le \tau \le 1.
\end{eqnarray}
It can be observed that using \emph{both} hierarchical modulation and time sharing enables to get some points which can not be obtained by using only one of the two methods.

\section{Performance Evaluation for Hierarchical Modulations}\label{part3}

We present in this section a method to get easily the spectrum efficiency and the required $E_s/N_0$ for a given targeted BER/PER for any modulation and coding rate without computing extensive simulations. In \cite{cnes}, a fast coding/decoding performance evaluation method based on the mutual information computation is proposed and applied to time varying channel for QPSK modulation with very good prediction precision. The method developped hererafter, based on channel capacity computation, makes it possible to predict the performances of a coding scheme combined with any modulation and especially any hierarchical modulation. The last part of this section presents the capacity region for real codes using our method.

\subsection{Principle}

\subsubsection{Ideal code}
Before applying the approach to a real code, we begin with an example using theoretical ideal codes achieving the channel capacity. The normalized capacity for a modulation is defined by $\overline{C}_{mod} = \frac{1}{m} C_{mod}$, where $C_{mod}$ is the modulation's capacity and m is the number of bits per symbol. The $\overline{C}_{mod}$ function belongs to $[0,1[$ and corresponds to the maximal coding rate of an ideal code, which achieves error-free transmission.

\begin{figure}[!ht]
\centering
\includegraphics[width = 0.5\textwidth]{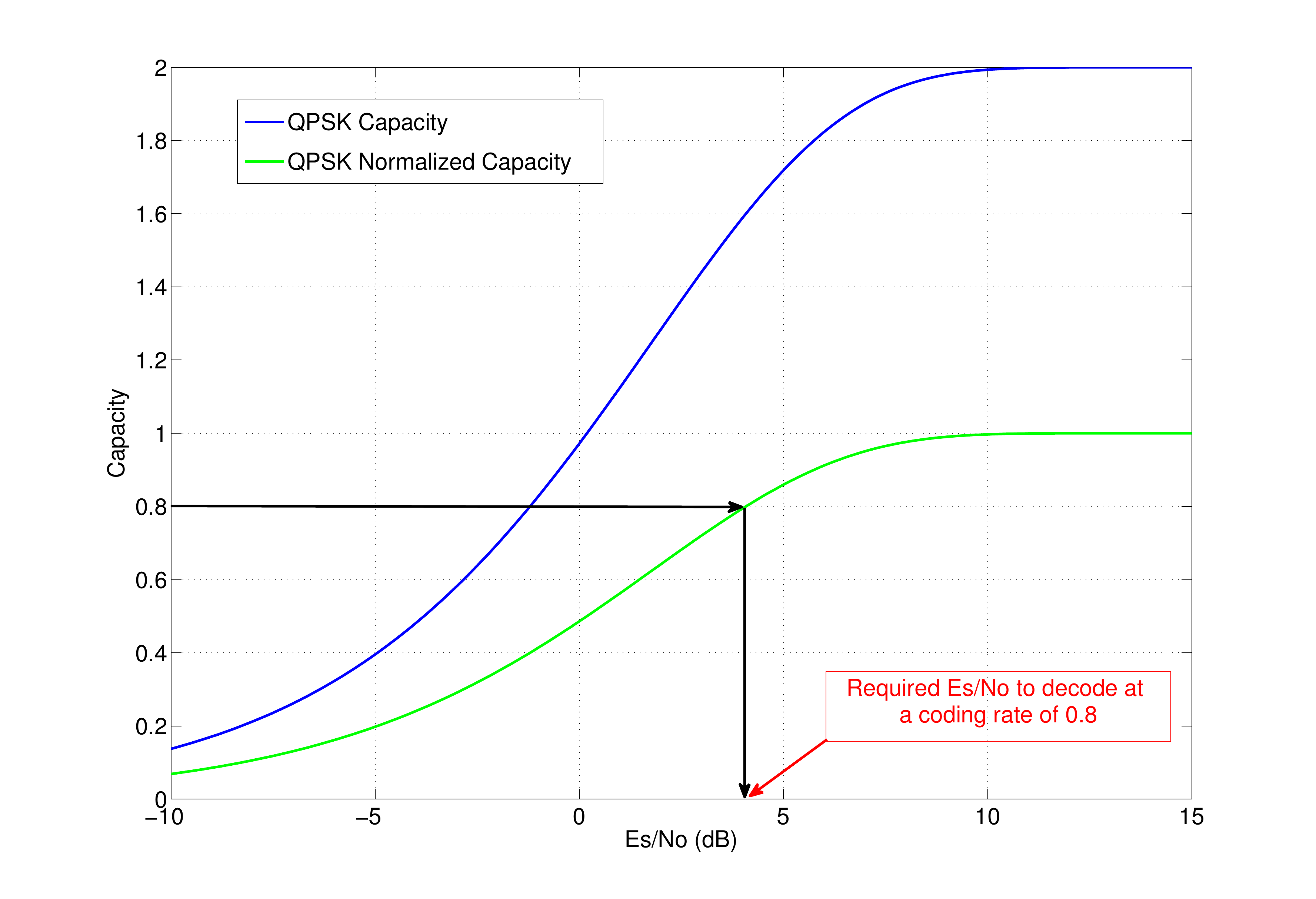}
\caption{Decoding threshold for an ideal code of rate 0.8}
\label{mean_capacity}
\end{figure}

Given a modulation and a coding rate, if we want to know at which $E_s/N_0$ the ideal code is able to decode, we just need to inverse the normalized capacity function for that coding rate. Figure~\ref{mean_capacity} illustrates this example using a QPSK modulation.

\subsubsection{Real code}
Let us now consider an actual waveform based on a real code. We can consider the real code with a coding rate R is similar to an ideal code with rate $\tilde R$ (in terms of decoding threshold), where $\tilde R \ge R$. To determine $\tilde R$, we propose to consider the performance curve (BER or PER vs $E_s/N_0$) for \emph{one actual reference modulation} and \emph{coding scheme with code rate $R$}. The principle consists in first determining from this performance curve a value $\left(E_s/N_0\right)_{ref}$ corresponding to a quasi error-free transmission (e.g., $\text{BER}=10^{-5}$). Then, we obtain $\tilde R$ as the value corresponding to $\left(E_s/N_0\right)_{ref}$ in the normalized capacity vs SNR curve of the reference modulation. Finally, we find the SNR operating point for the hierarchical modulation by taking the value corresponding to $\tilde R$ in the normalized capacity vs SNR curve of the hierarchical modulation. The set of operating points for different code rates gives the spectrum efficiency curve of the hierarchcial modulation. We illustrate our method with an example. We would like to study the SNR operating points of the hierarchical modulation using an non-uniform 16-QAM ($\alpha=2$) in the DVB-SH standard at a target BER of $10^{-5}$. We use the 2/9 and 1/5-turbo codes for the HP and LP streams respectively. To determine $\tilde R$ and the decoding thresholds, the method works as follow:

\begin{enumerate}
\item Use the performance curve of the reference modulation with rate R to get the operating point $\left(E_s/N_0\right)_{ref}$ such as we have the desired performance. In the DVB-SH guidelines \cite[Table 7.5]{sh}, we read: 
\begin{eqnarray*}
\text{For the coding rate 2/9: }\text{BER}_{\text{QPSK}} \left( -3.4 \text{dB} \right) = 10^{-5} \Rightarrow \left(E_s/N_0\right)_{ref} = -3.4 \text{dB}\\
\text{For the coding rate 1/5: }\text{BER}_{\text{QPSK}} \left( -3.9 \text{dB} \right) = 10^{-5} \Rightarrow \left(E_s/N_0\right)_{ref} = -3.9 \text{dB}
\end{eqnarray*}

\item Compute the normalized capacity for the reference modulation, which corresponds to $\tilde R$ (see Figure~\ref{mean_capacity}): 
\begin{eqnarray*}
\text{For the HP stream: }\tilde R &=& \overline{C}_{\text{QPSK}} \left( -3.4 \text{dB} \right) \thickapprox 0.27\\
\text{For the LP stream: }\tilde R &=& \overline{C}_{\text{QPSK}} \left( -3.9 \text{dB} \right) \thickapprox 0.2455
\end{eqnarray*} 

\item For the studied modulation, compute $E_s/N_0$ such as the normalized capacity at this SNR equals $\tilde R$:
\begin{eqnarray*}
\left(E_s/N_0\right)_{HP} &=& \overline{C}_{\text{HP,}\alpha=2}^{-1} \left( \tilde R = 0.27 \right) = -2.7\text{dB}\\
\left(E_s/N_0\right)_{LP} &=& \overline{C}_{\text{LP,}\alpha=2}^{-1} \left( \tilde R = 0.2455 \right) = 6.2\text{dB}
\end{eqnarray*}
The DVB-SH guidelines give the decoding thresholds for the HP and LP streams \cite[Table 7.40]{sh}. We read -2.6dB and 6.5dB for the HP and LP streams respectively (we remove the 0.3dB due to the pilots).

\item Finally the points $\left( \left(E_s/N_0\right)_{HP}, R_{HP} \times m \right)$ and $\left( \left(E_s/N_0\right)_{LP}, R_{LP} \times m \right)$ are plotted on the spectrum efficiency curve (e.g. Figure~\ref{eff_spec}).
\end{enumerate}

These steps are repeated for all the coding rates. Our method makes two assumptions. First of all, it approximates the information rate by $R \times m$. This approximation is justified by the fact that the targeted performance ($\text{BER}=10^{-5}$) is very small and thus only hardly impact the useful information rate.

\begin{figure}[!ht]
\centering
\includegraphics[width = 0.5\textwidth]{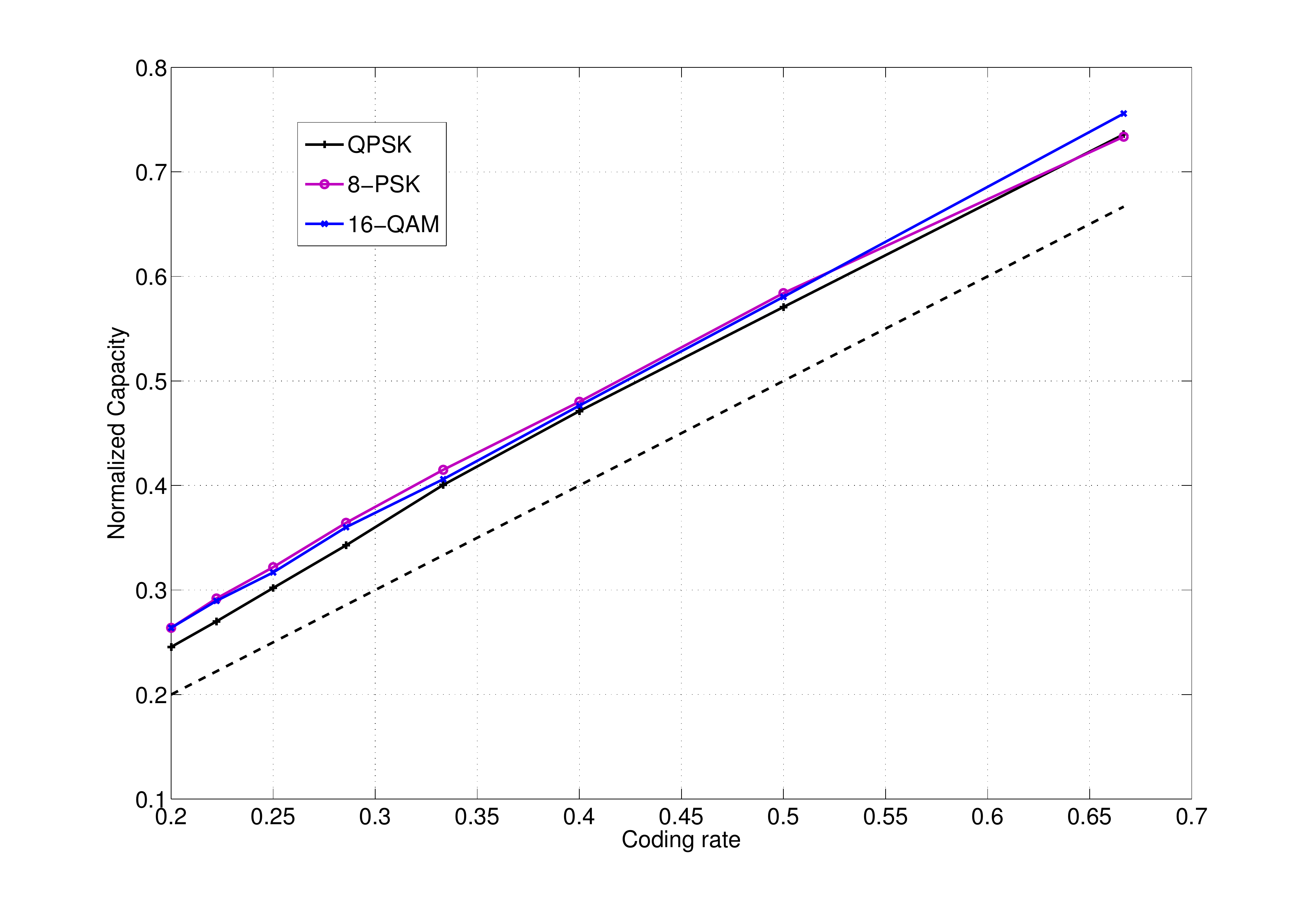}
\caption{DVB-SH: normalized capacities}
\label{mean_capacity_vs_coding}
\end{figure}

The second assumption is to suppose as in \cite{cnes} that \emph{the performance of the decoding only depends on the normalized capacity and not on the modulation as for ideal codes}. To validate this second assumption, we present on Figure~\ref{mean_capacity_vs_coding} the normalized capacity in function of the coding rate for various modulations using the data of DVB-SH \cite[Table 7.5]{sh}. For ideal codes, the points are merged and located on the dotted line. But for real codes, we can remark that, even if the points are different, they are closed.

\subsection{Application to DVB-SH}
In our study, we use the guidelines of DVB-SH \cite[Table 7.5]{sh} to get the reference curves and the operating points at a target BER of $10^{-5}$. These data correspond to a static receiver. The guidelines also provide all the reference numerical results for the hierarchical modulation \cite[Figure 7.40]{sh}, where 0.3dB need to be removed due to the pilots. These results allow to evaluate the efficiency of our method.

\begin{figure*}[!ht]
\centerline{\subfloat[$\alpha=2$]{\includegraphics[width=0.5\columnwidth]{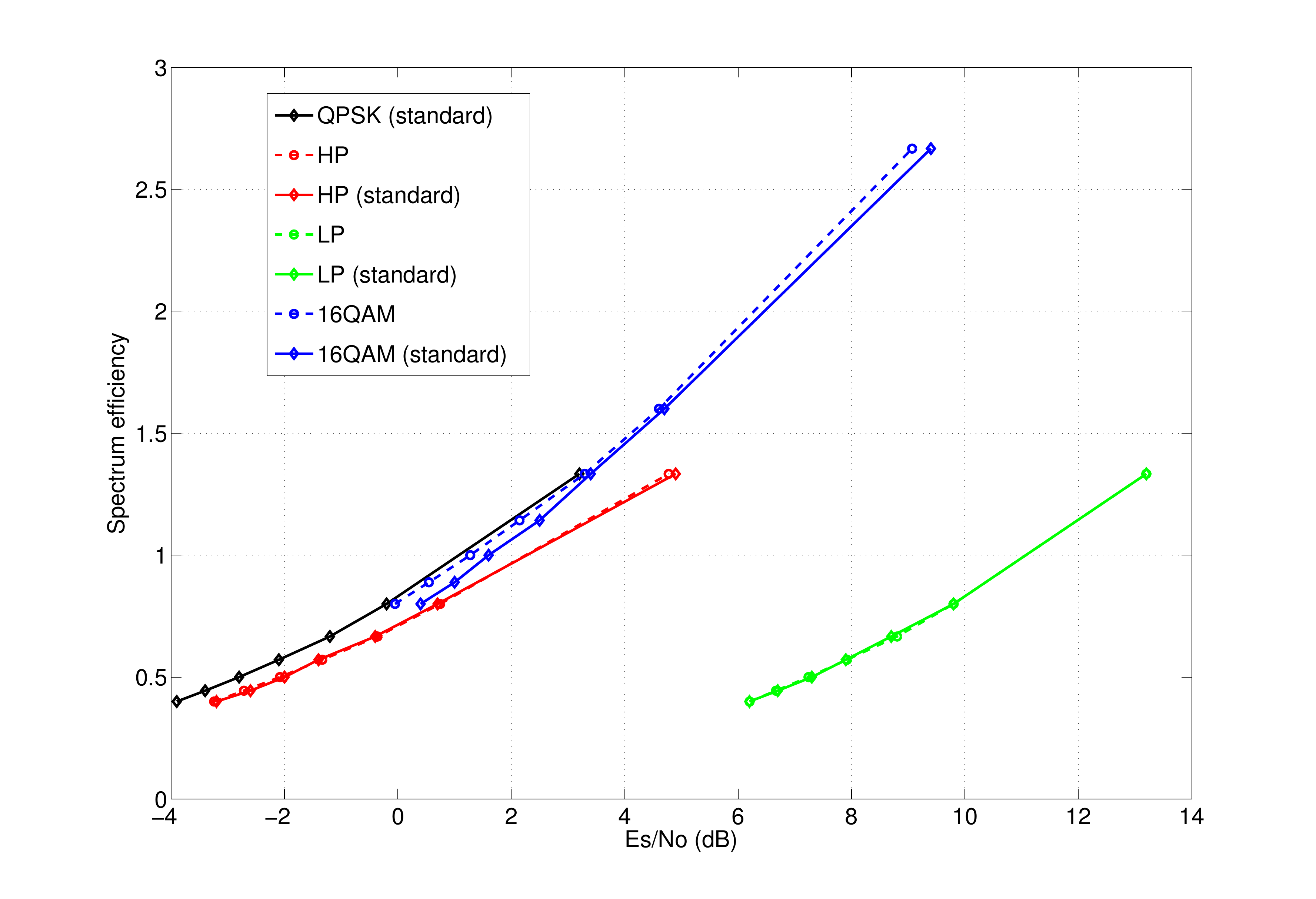}%
\label{eff_spec_2}}%
\hfil
\subfloat[$\alpha=4$]{\includegraphics[width=0.5\columnwidth]{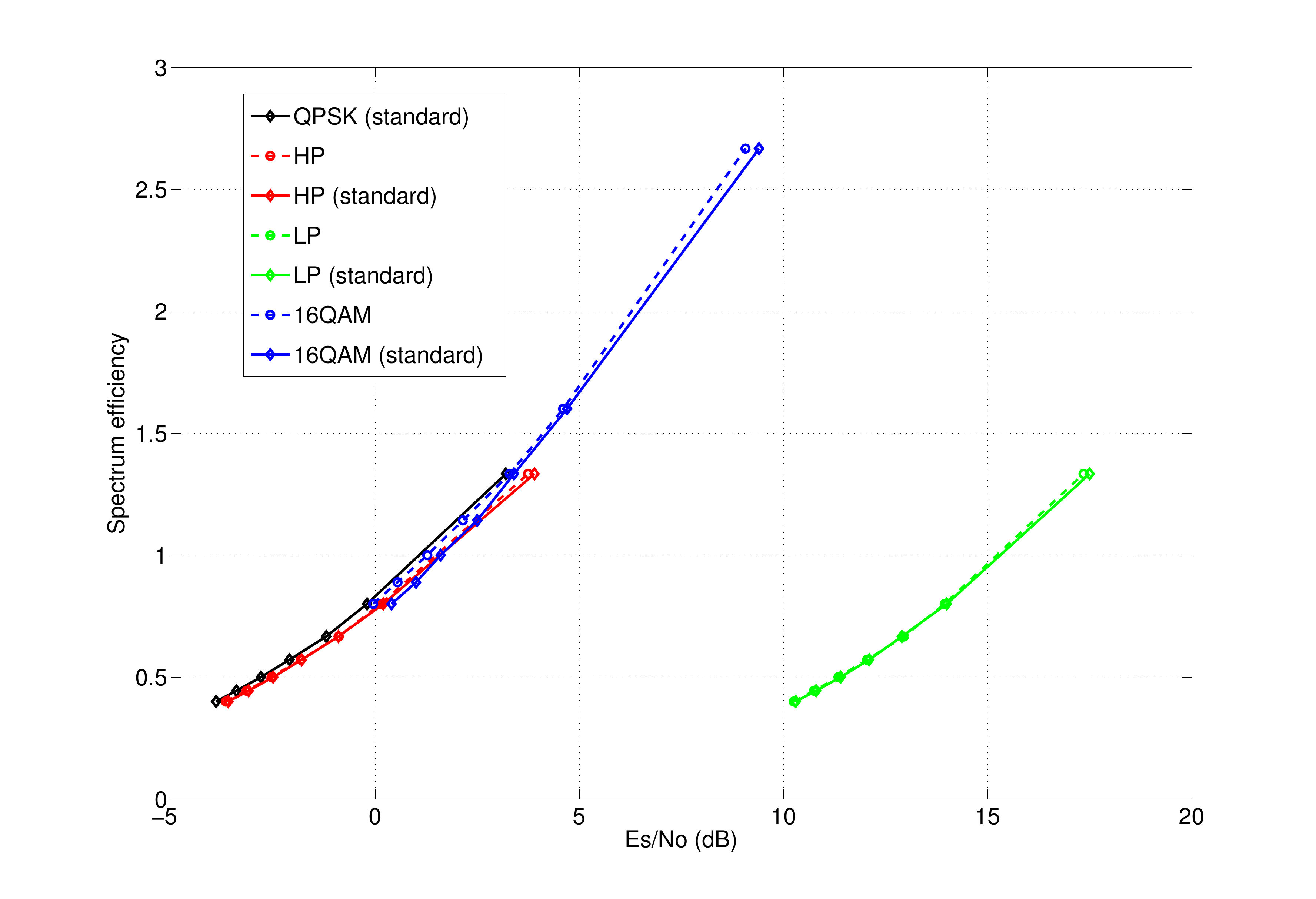}%
\label{eff_spec_4}}}%
\caption{DVB-SH spectrum efficiency, $\text{BER}=10^{-5}$}
\label{eff_spec}
\end{figure*}

The method described earlier is now applied to plot the spectrum efficiency curves as a function of required $E_s/N_0$ (target $\text{BER}=10^{-5}$). DVB-SH considers two values of $\alpha$, 2 and 4. The curves are given in Figure~\ref{eff_spec}. The reference results from the guidelines correspond to the standard curves. The results show a good precision and it does not require computing extensive simulations. Moreover, our work consolidates the fact that the capacity is a good metric to evaluate performance.

\subsection{DVB-SH: Additional Results}
We apply hereafter the method to compute the required $E_s/N_0$ in function of $\alpha$ or the coding rate.

\begin{figure*}[!ht]
\centerline{\subfloat[HP Stream]{\includegraphics[width=0.5\columnwidth]{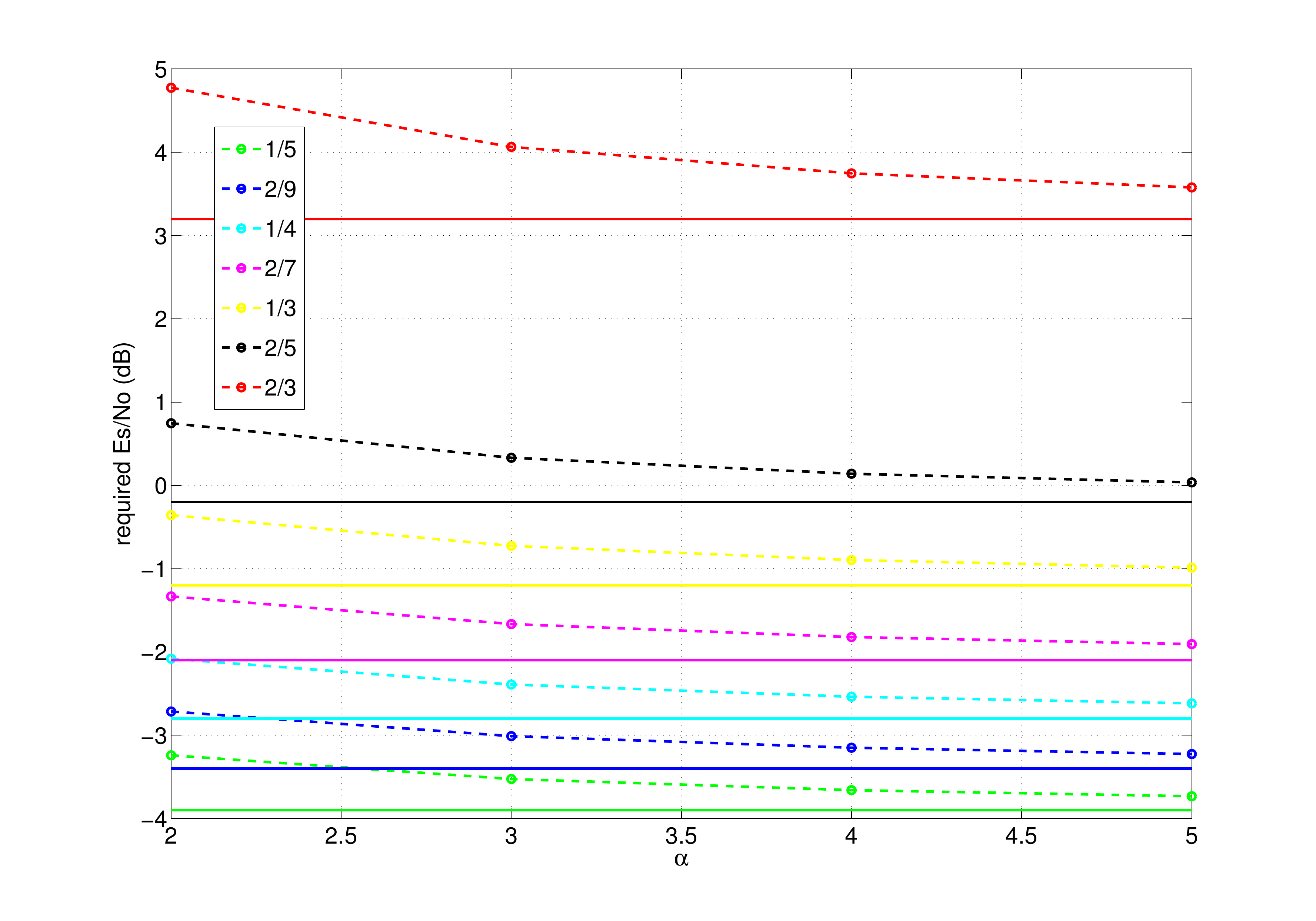}%
\label{sgn_vs_alpha_hp}}%
\hfil
\subfloat[LP Stream]{\includegraphics[width=0.5\columnwidth]{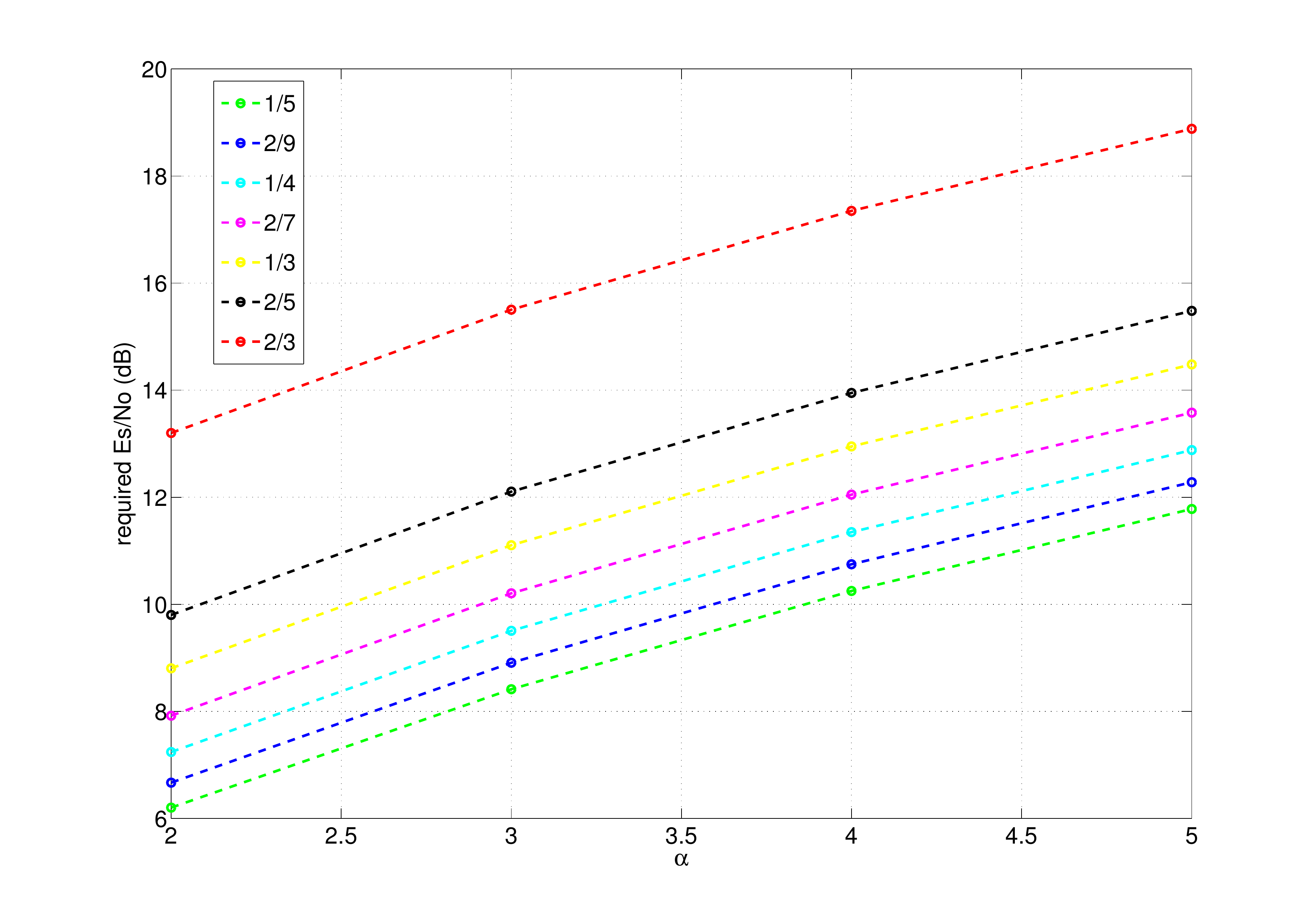}%
\label{sgn_vs_alpha_lp}}}%
\caption{Required $E_s/N_0$ function of $\alpha$, $\text{BER}=10^{-5}$}
\label{sgn_vs_alpha}
\end{figure*}

Figure~\ref{sgn_vs_alpha} presents how the required $E_s/N_0$ varies with $\alpha$. For the LP stream, the required $E_s/N_0$ is an increasing function of $\alpha$, unlike the one for the HP stream who decreases. In fact, the required $E_s/N_0$ for the HP stream tends to the required SNR of the QPSK modulation (continuous lines on Figure~\ref{sgn_vs_alpha_hp}). These results are obvious when we look the modification of the constellation with $\alpha$. When $\alpha$ increases, the points in one quadrant become closer and the constellation is similar to a QPSK. It explains why the LP stream is harder to decode and requires a better SNR.

\begin{figure*}[!ht]
\centerline{\subfloat[Required $E_s/N_0$ function of the coding rate, $\alpha=2$]{\includegraphics[width=0.5\columnwidth]{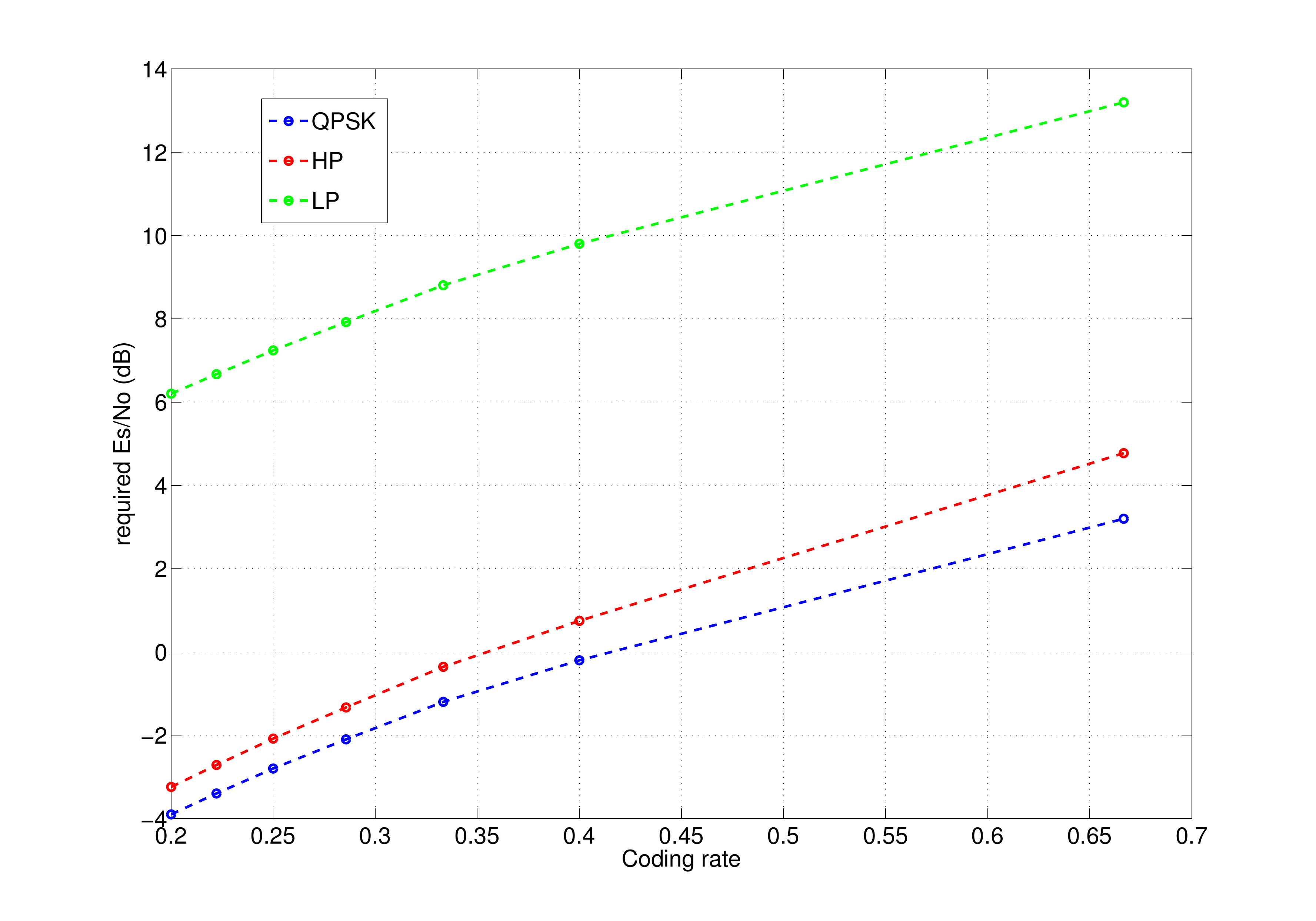}%
\label{sgn_vs_coding_2}}%
\hfil
\subfloat[Difference between the required $E_s/N_0$ for the QPSK and the HP/LP streams, $\alpha=2$]{\includegraphics[width=0.5\columnwidth]{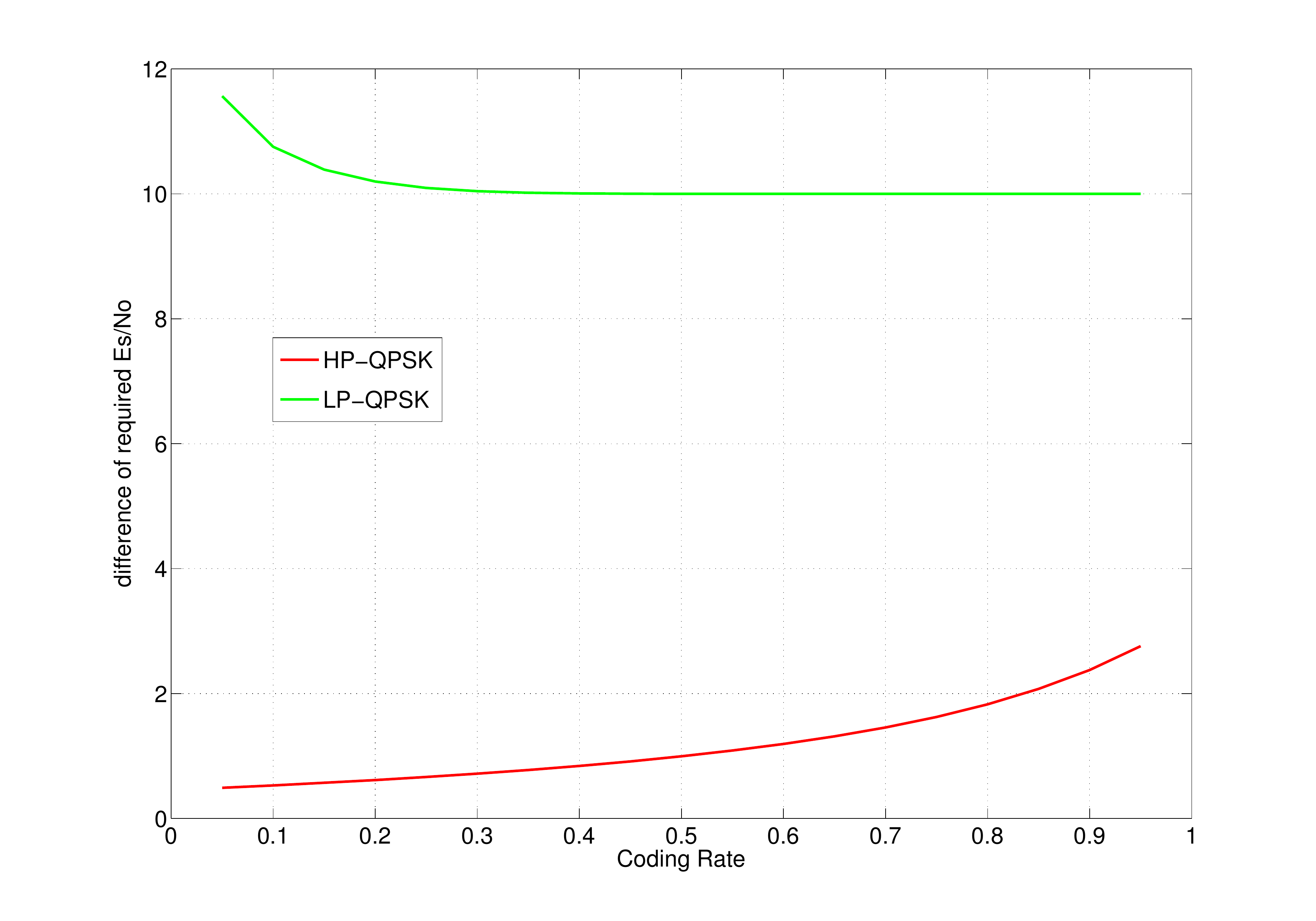}%
\label{delta_2}}}%
\caption{Study of the required $E_s/N_0$ function of the coding rate, $\text{BER}=10^{-5}$}
\label{sgn_vs_coding}
\end{figure*}

The last result concerns the variations of the required $E_s/N_0$ with the coding rate. Figure~\ref{sgn_vs_coding_2} shows the result for $\alpha=2$. Obviously if there is less redundancy, the SNR has to be higher to decode. Figure~\ref{delta_2} presents the difference between the required SNR for the QPSK and the HP/LP streams function of the coding rate with $\alpha=2$. We see that, for a coding rate $R \ge 0.3$, the difference between the QPSK and the LP stream is constant, which has been observed and quantified in \cite{tas}.

\subsection{Application to DVB-S2}
\begin{figure*}[!ht]
\centerline{\subfloat[$\theta=10^\circ$]{\includegraphics[width=0.5\columnwidth]{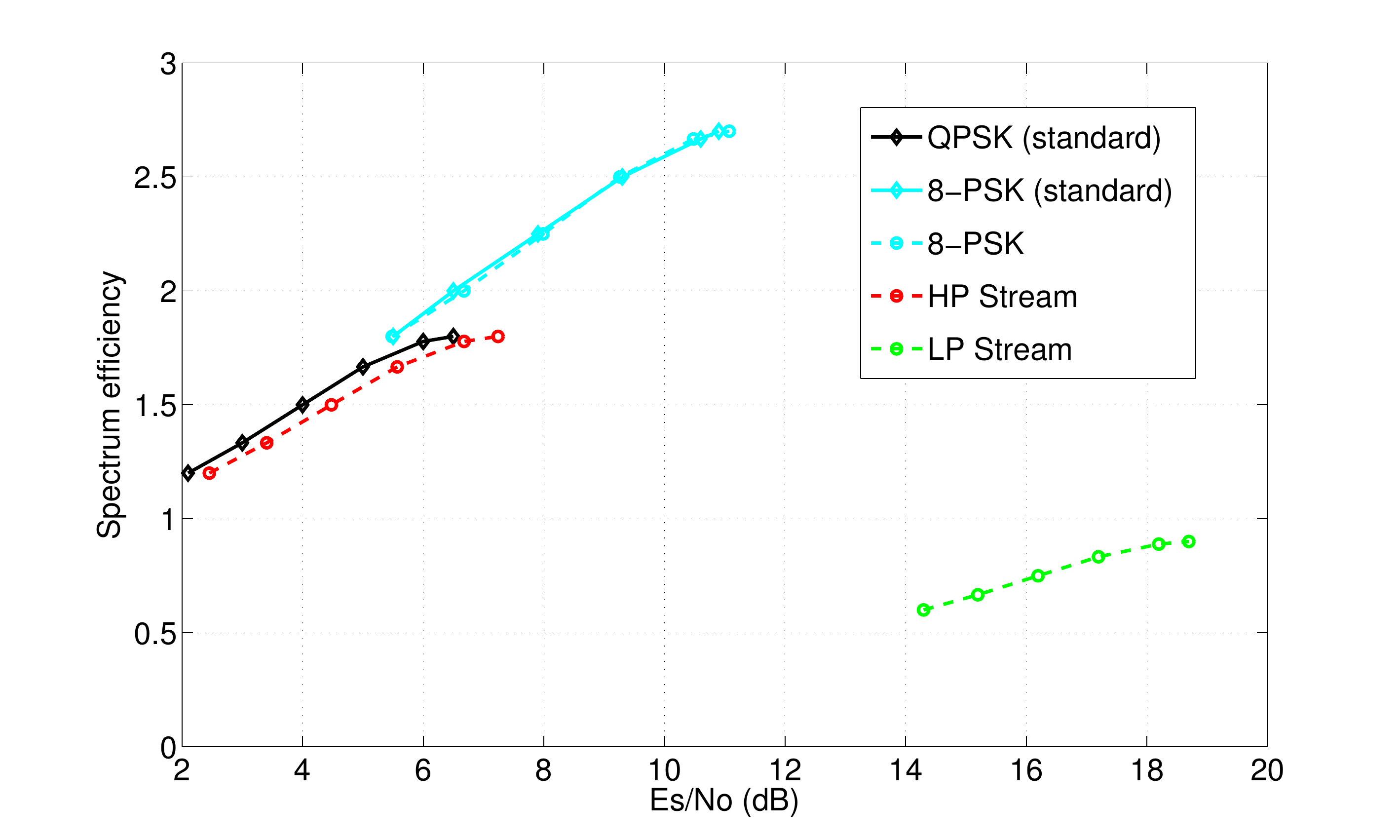}%
\label{eff_spec_10}}%
\hfil
\subfloat[$\theta=15^\circ$]{\includegraphics[width=0.5\columnwidth]{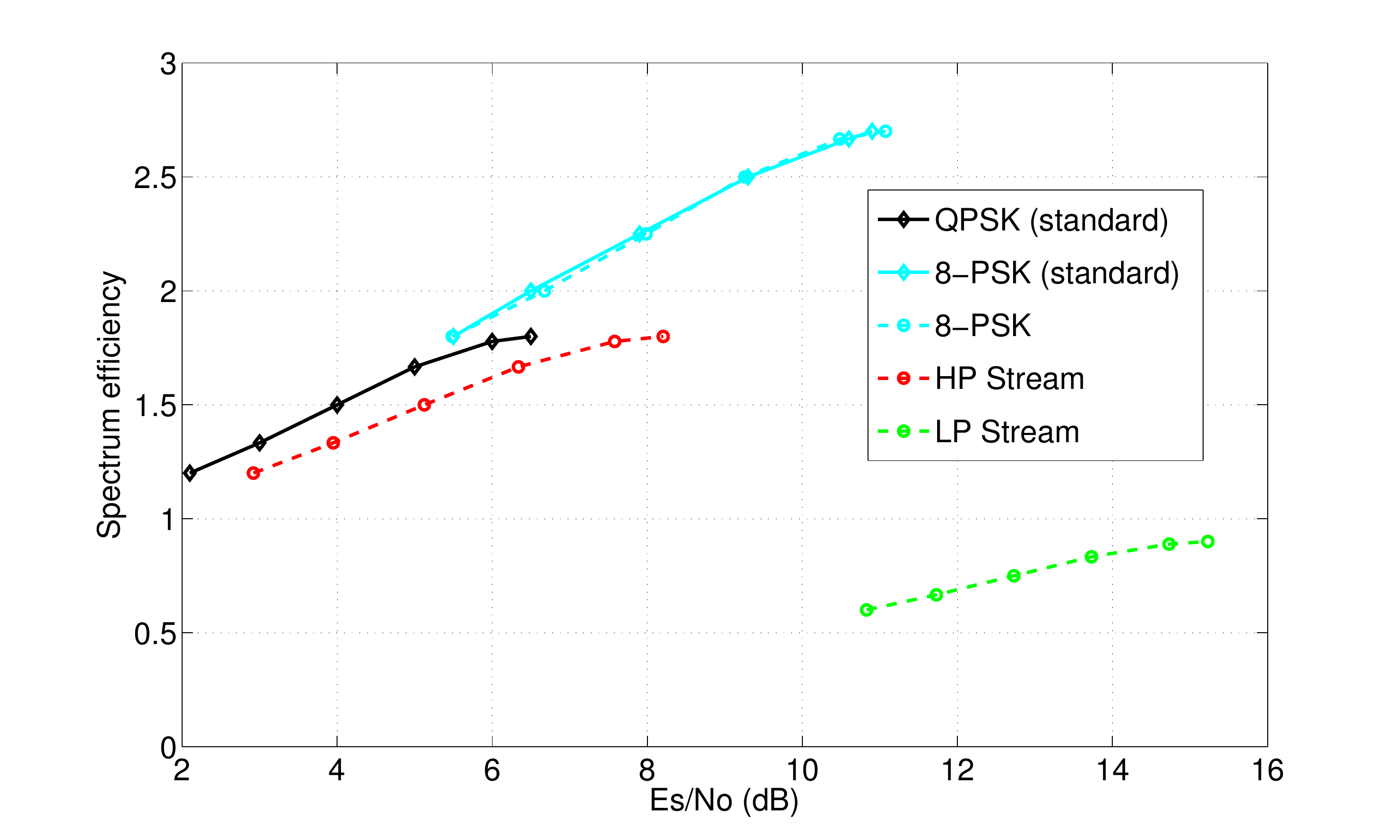}%
\label{eff_spec_15}}}%
\caption{DVB-S2 spectrum efficiency, $\text{PER}=10^{-5}$}
\label{eff_spec_s2}
\end{figure*}

The performance curves for the LDPC codes used in DVB-S2 are given in \cite[Fig~7]{ldpc_s2}. The spectrum efficiency and the required $E_s/N_0$ can be computed using the method described previously. Here we choose a target PER of $10^{-5}$ as desired performance. Figure~\ref{eff_spec_s2} presents the results for two values of $\theta$. The standard does not explicit any value for $\theta$ and reference numerical results. To lighten the paper, we do not present any additional curve to Figure~\ref{eff_spec_s2} as in the previous part. In conclusion to this section, we can remark that $\theta$ plays the same role as $\alpha$: it allows to modify the performance of each stream.
 
\subsection{Time Sharing vs Hierarchical Modulation}
We are interested here to obtain the equivalent of the capacity region for real codes. This can be considered as spectrum efficiency region.

As described in Section~\ref{part2}, the available energy is shared between two flows for the superposition coding strategy. We need to compute the achievable rate for the real code. Once the power allocation is done and $\alpha$ determined, we read on the spectrum efficiency curve the capacity for each population. Unfortunately, the spectrum efficiency curve is not a continuous fonction since it has been computed for the different coding rates allowed by the standard. To obtain the capacity for any SNR, we approximate the spectrum efficiency by linear interpolation between each point as presented in Figure~\ref{eff_spec_s2}. We reminder the achievable rates for each population, where $SNR_1 \le SNR_2$,
\begin{eqnarray}
 R_1 &=& C_{hp} \left( SNR_1 \right) , \nonumber\\
 R_2 &=& C_{hp}\left( SNR_1 \right) + C_{lp}\left( SNR_2 \right).
\label{achievable_rates}
\end{eqnarray}

\begin{figure*}[!ht]
\centerline{\subfloat[DVB-SH: SNR$_1$=2dB, SNR$_2$=10dB]{\includegraphics[width=0.5\columnwidth]{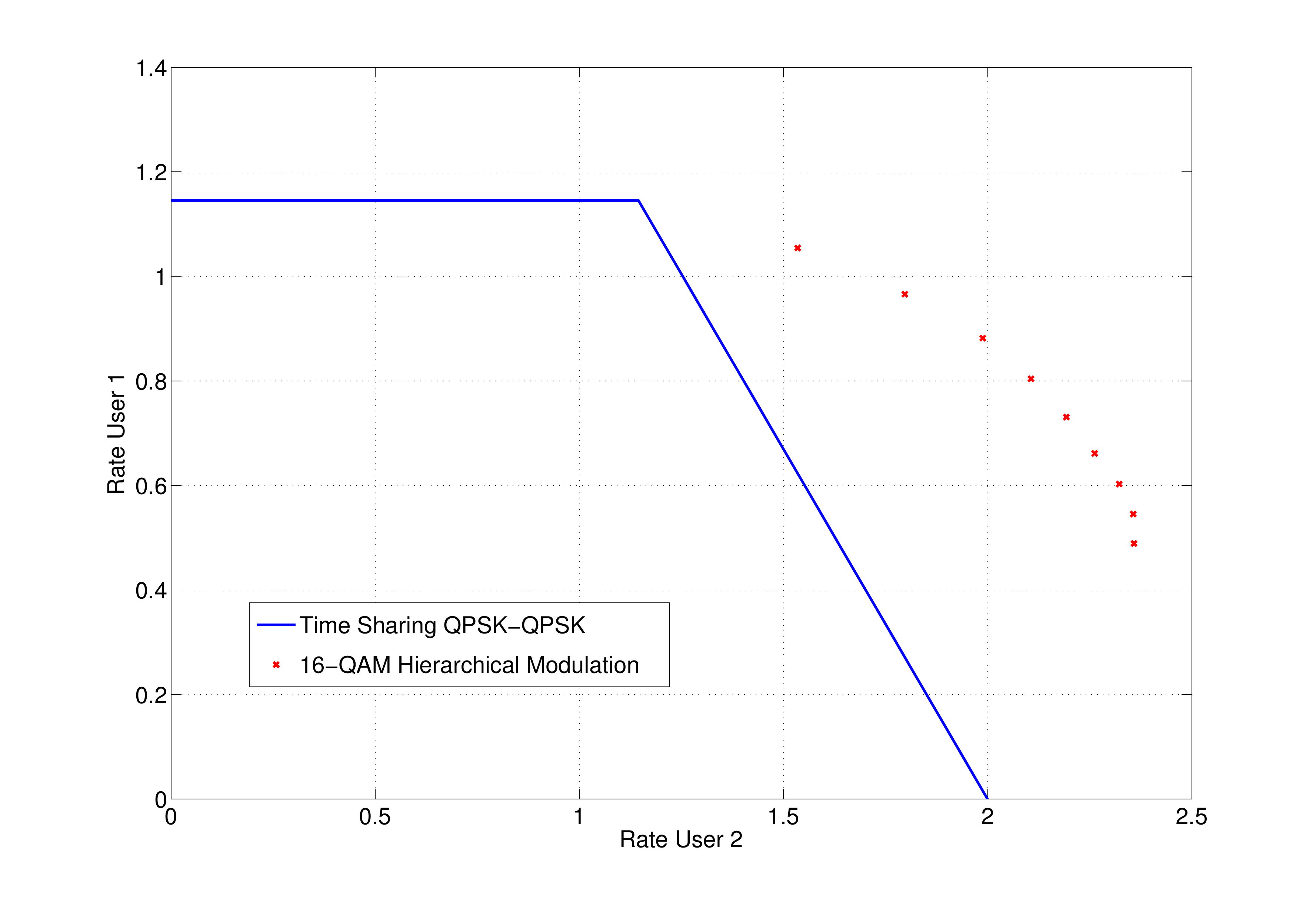}%
\label{real_2_10}}%
\hfil
\subfloat[DVB-S2: SNR$_1$=9dB, SNR$_2$=10dB]{\includegraphics[width=0.5\columnwidth]{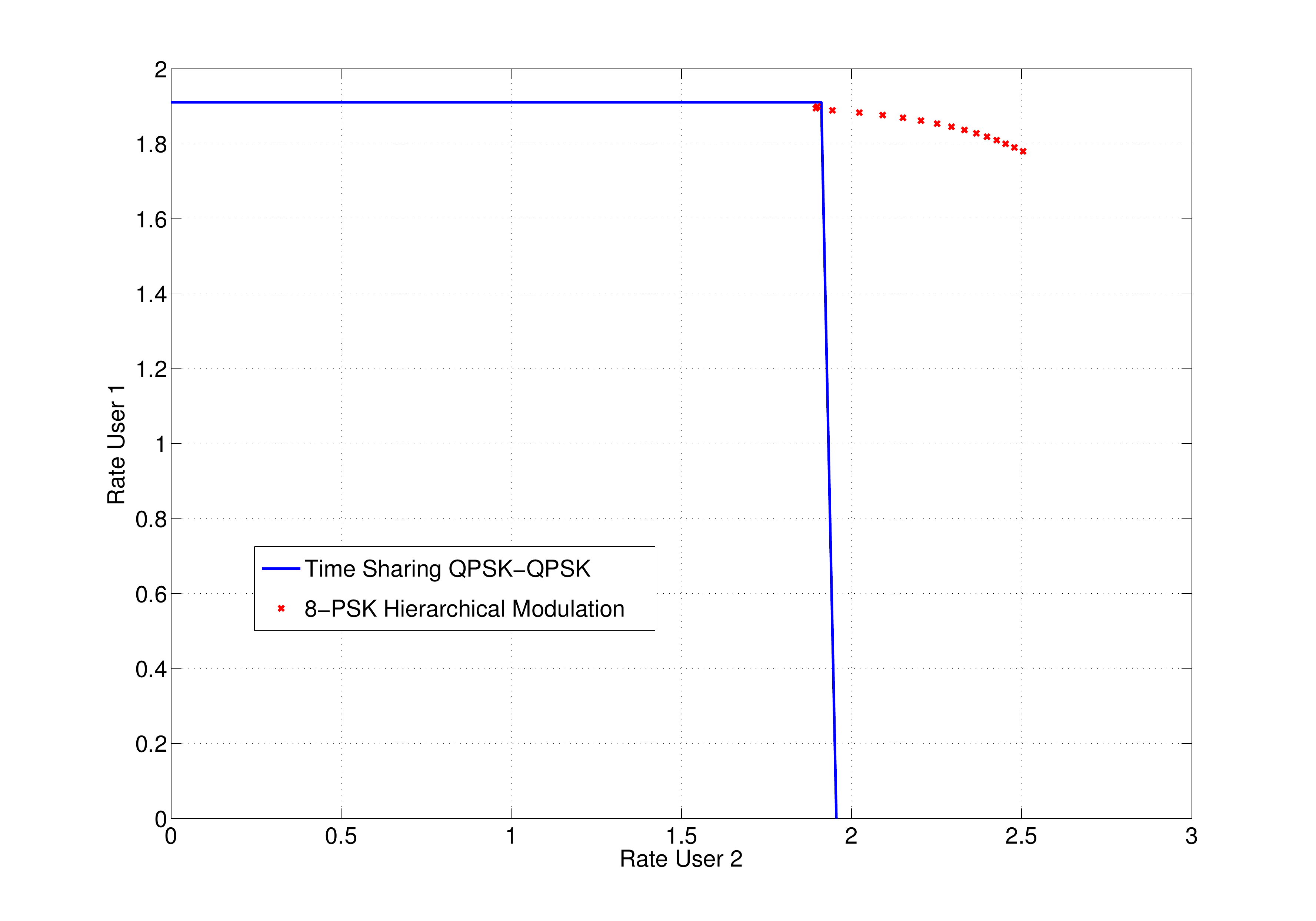}%
\label{real_9_10}}}%
\caption{Capacity Region}
\label{real_capacity_region}
\end{figure*}

Figure~\ref{real_capacity_region} presents the results using this interpolation for DVB-SH and DVB-S2. Here again the hierarchical modulation outperforms the time sharing strategy with two QPSK modulations for several SNR configurations (we choose the different SNR according to distributions given in Section~\ref{part4}). If the time sharing uses different modulations than QPSK, it is also possible to represent the achievable rates. Theses curves are particurlarly interesting when using adaptive modulation as in DVB-S2 \cite{s2}. In that case, it is possible to identify for a  given SNR configuration, which solution is the best: hierarchical modulation or time sharing (with all possible modulations).

\section{Spectrum efficiency vs Indisponibility}\label{part4}

In this section, we study a broadcast system, where the receivers SNR distribution is known. We define the indisponibility and compare hierarchical modulation to classical modulations using this new criteria in addition to the spectrum efficiency.

\subsection{Definition of Spectrum Efficiency and Indisponibility}
In the previous part, we introduce a method based on the capacity in order to estimate the spectrum efficiency for any modulation and in particular hierarchical one combined with real codes. It is then possible to compare various modulations in terms of spectrum efficiency and power performance. The next step is to take into account the channel variability in the dimensionning of a broadcast system. The indisponibility is in that case relevant to complete the spectrum efficiency criteria in the choice of modulation and coding scheme. The indisponibility is simply defined as the percent of the population which can not decode any stream. Its computation requires SNR distributions. This notion completes the spectrum efficiency in the sense that the coding scheme maximising the spectrum efficiency may also be decoded by a small fraction of the population, which is not admissible. A compromise has to be found between a good spectrum efficiency and a tolerable indisponibility. We consider here a mean spectrum efficiency over the population who receive at least the HP stream. Equation~(\ref{mean_spectrum_efficiency}) gives the mean spectrum efficiency formula, where $\mu_{x}$ represents the spectrum efficiency for the stream $x$ and $\rho_{x}$ is the percent of the population decoding the stream $x$ with the inequality $\rho_{lp} \le \rho_{hp}$.\\
\begin{eqnarray}
\text{Mean Spectrum Efficiency} = \frac{ \mu_{hp}\rho_{hp}+ \mu_{lp}\rho_{lp}}{\rho_{hp}}
\label{mean_spectrum_efficiency}
\end{eqnarray}

For instance in the best case, all the population decode both streams so $\rho_{hp}=\rho_{lp}=1$ and the mean spectrum efficiency equals $\mu_{hp}+\mu_{lp}$.

\subsection{Application to DVB-SH}
\begin{figure*}[!ht]
\centerline{\subfloat[SNR distribution for several environments]{\includegraphics[width=0.5\columnwidth]{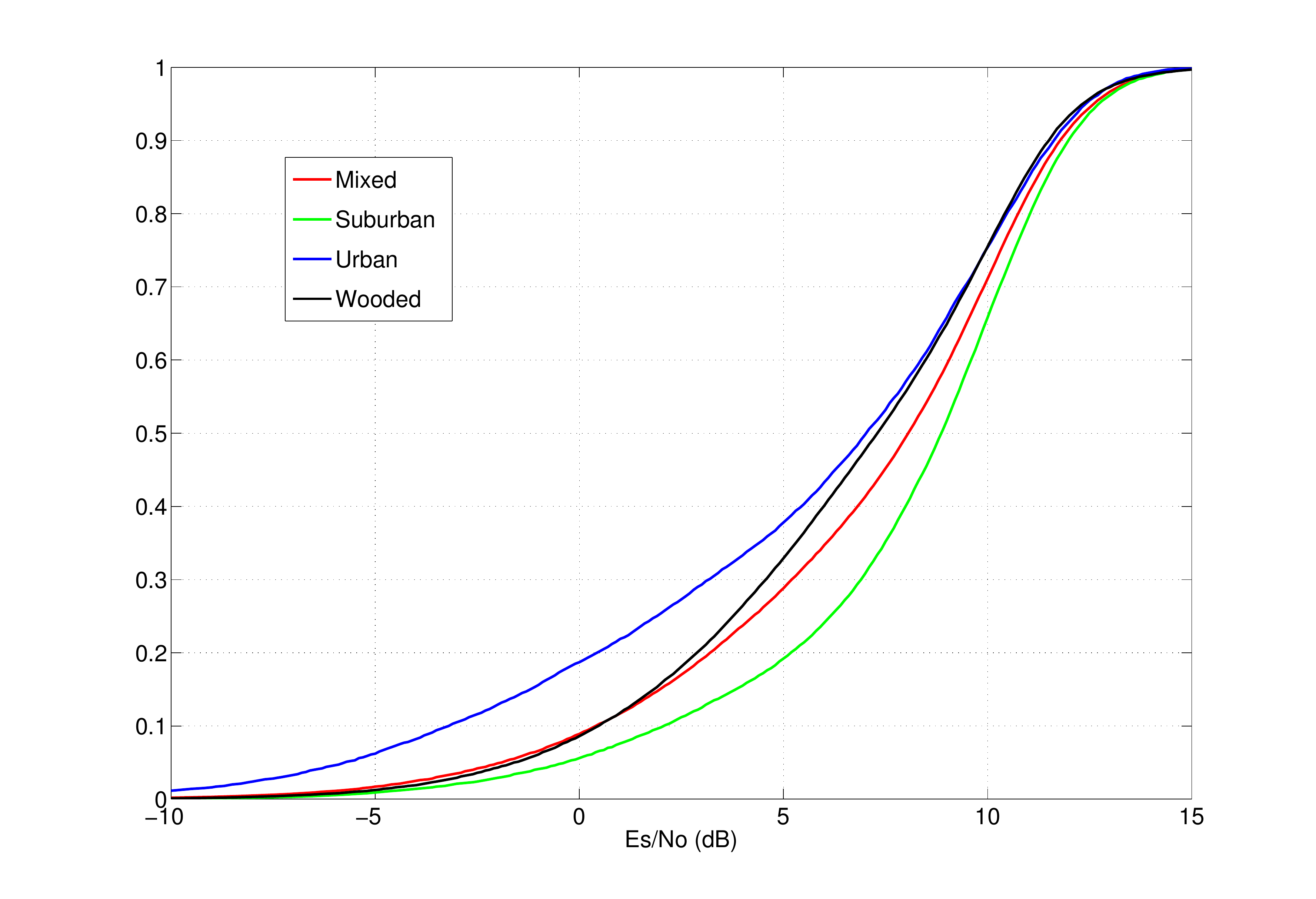}%
\label{S_distrib}}%
\hfil
\subfloat[Indisponibility vs spectrum efficiency: mixed environment]{\includegraphics[width=0.5\columnwidth]{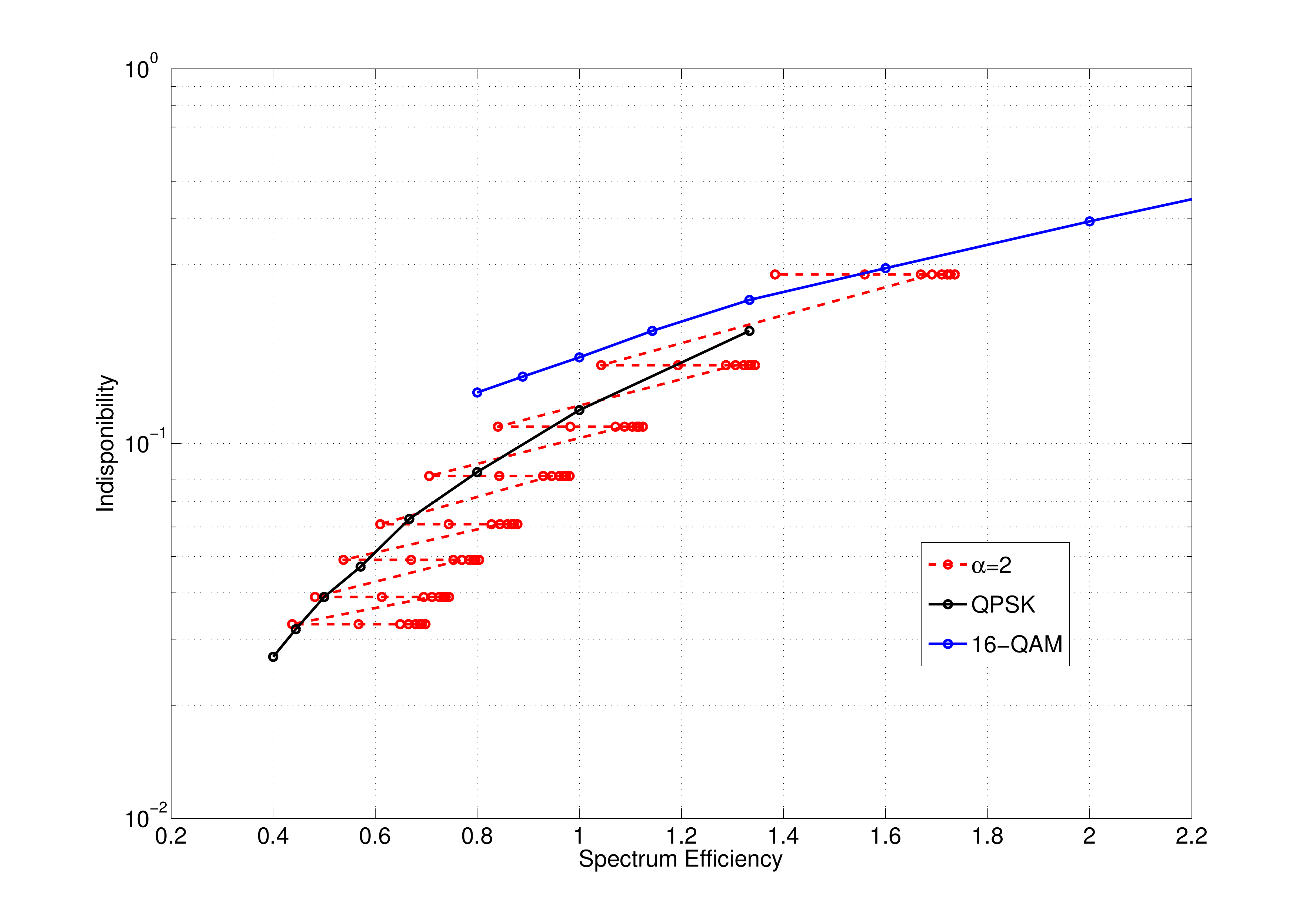}%
\label{environment}}}%
\caption{DVB-SH}
\label{indispo_sh}
\end{figure*}

Figure~\ref{S_distrib} presents the SNR distribution for several environments in S-band due to shadowing and fading effects of these environments. The distribution is the result of measures realized by the CNES in 2008. The foreseen application is here satellite multimedia broadcasting to handheld mobile terminals.

Using these distributions, it is possible to compute the indisponibility for any configuration of the hierarchical modulation. Figure~\ref{environment} presents the results for one environment. For the hierarchical modulation, once the constellation parameter and the coding rate of the HP stream have been set, it is possible to compute the indisponibility who only depends on the required SNR to decode the HP stream. For a given HP coding rate, we choose to represent on Figure~\ref{environment} the points which verify the constraint $\frac{E_s}{N_0}_{lp} \ge \frac{E_s}{N_0}_{hp}$.

The coding rate of the LP stream has an impact on the ponderate spectrum efficiency but not on the indisponibility. An interesting fact is when the coding rate of the LP stream for a given HP coding rate is increased, it is expected that the ponderate spectrum efficiency increases but it is not always the case. It can be explained by the computation of the ponderate spectrum efficiency in (\ref{mean_spectrum_efficiency}). When the LP coding rate grows, $\mu_{lp}$ increase but in the same time $\rho_{lp}$ decrease. As $\rho_{lp}$ depend on the SNR distribution, the environment has an impact on the result.

We now focus on Figure~\ref{environment}. We are interested to find the best configuration for an indisponibility around $10^{-1}$ (10\%). The 16-QAM is not an option as it does not reach that level of indisponibility taking into account the DVB-SH available code rates. The best spectrum efficiency is achieved using a hierarchical modulation with $\alpha=2$, coding rates 2/5 and 1/5 for the HP and LP streams respectively. In that particular case, the hierarchical modulation offers better performance than classical modulations such as the QPSK and the 16-QAM. Note that  there is no general rule to choose the good modulation/coding scheme as the SNR distribution has a great impact on the final choice. However, in many cases, it appears that hierarchical modulation is significantly better than classical modulations.

Finally, the principle can be applied to more complicated scenario, where all the population does not experience the same environment.

\subsection{Application to DVB-S2} 
\begin{figure*}[!ht]
\centerline{\subfloat[SNR distribution]{\includegraphics[width=0.5\columnwidth]{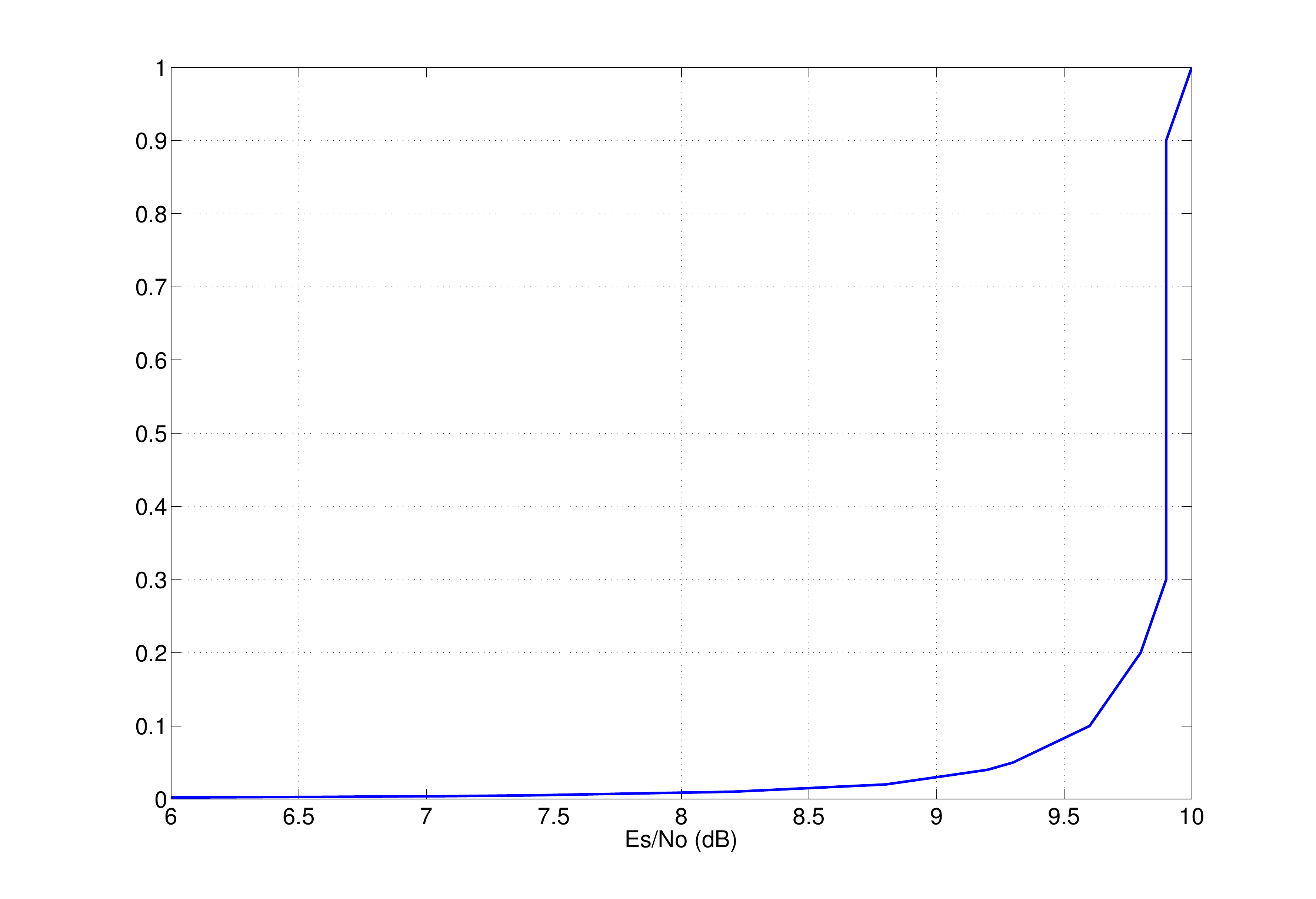}%
\label{ka_distrib}}%
\hfil
\subfloat[Indisponibility vs spectrum efficiency: $\theta=10^\circ$]{\includegraphics[width=0.5\columnwidth]{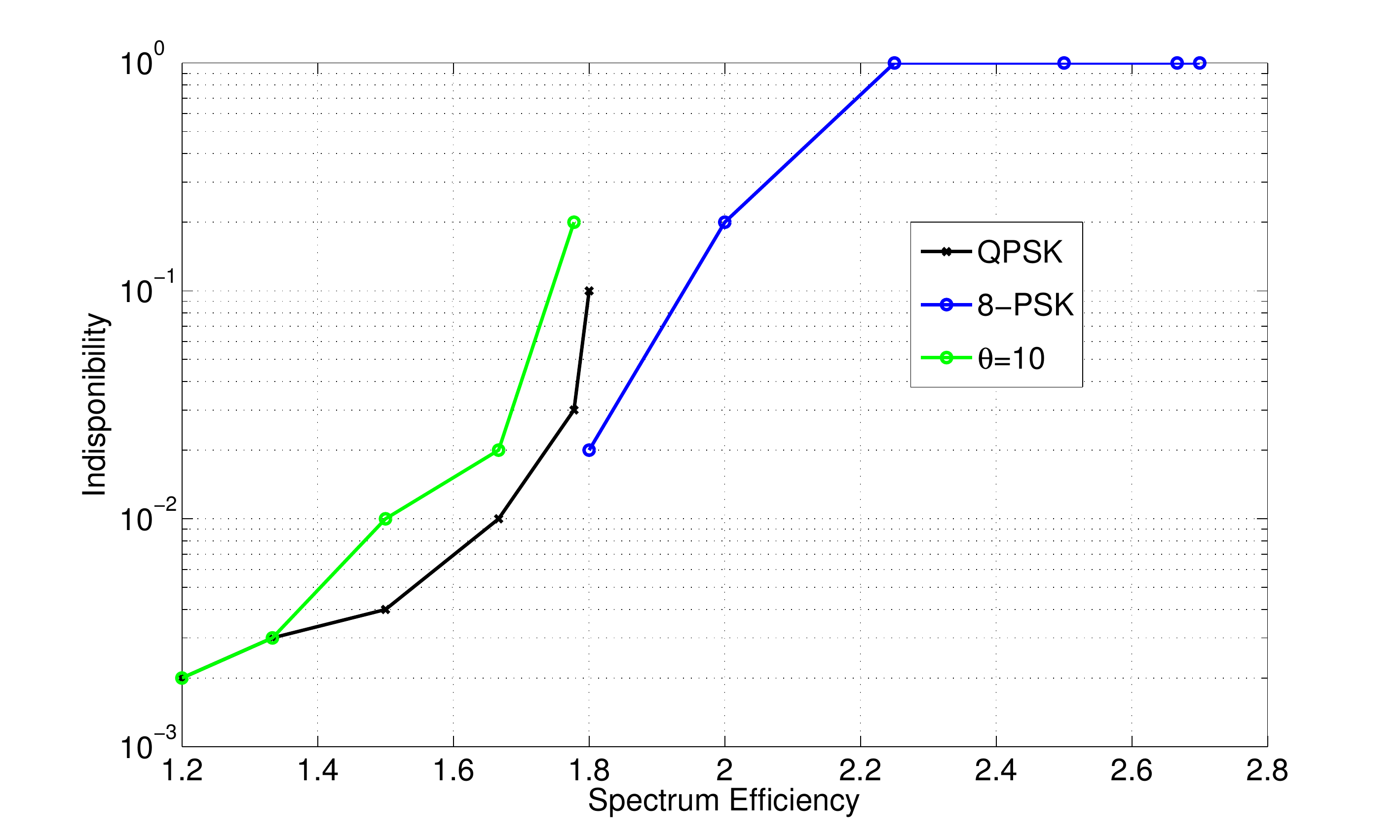}%
\label{environment_s2}}}%
\caption{DVB-S2}
\label{indispo_s2}
\end{figure*}

The SNR distribution for DVB-S2 on Figure\ref{ka_distrib} represents the fading in the Ka band due to rain attenuation. It is quite different from the distributions presented on Figure~\ref{S_distrib}. The slope of the curve is much steeper in that case.

This also has an impact on the performance of the hierarchical modulation. The hierarchical 8-PSK presents here a lower spectrum efficiency compared to the QPSK and for a given indisponibility. Figure~\ref{environment_s2} presents the results for $\theta=10^\circ$. In that case, the hierarchical modulation does not outperform the 8-PSK and QPSK modulations. However, compared to the 8-PSK, it is possible to obtain lower indisponibility due to the impact of $\theta$. 

\section{Conclusion}\label{conclusion}

In this paper, we introduce a general method allowing to analyse the performance of hierarchical modulations. This method relies on the channel capacity, which has been computed for any kind of constellation. It has been first applied to DVB-SH and DVB-S2 in order to obtain the spectrum efficiency. Comparisons with reference numerical results show the good reliability of the method.

We also compare the performance of hierarchical modulation with time sharing in terms of achievable rates. The results show that hierarchical modulation often outperform the other scheme.

To go further we introduce the notion of indisponibility using SNR distributions. It permits to compare different modulations with two criteria: mean spectrum efficiency and indisponibity. The result is useful in order to pick up a modulation when dimensionning a broadcast system and is illustrated on two application examples: DVB-SH S-band broadcasting to mobile handheld terminals and DVB-S2 Ka-band broadcasting to fixed terminals.

\section*{Acknowledgment}
The authors wish to thank Fr\'ed\'eric Lacoste for sharing the SNR distributions presented in Section~\ref{part4}.

\ifCLASSOPTIONcaptionsoff
  \newpage
\fi



%

\nocite{*}
\bibliographystyle{IEEEtran}
\bibliography{biblio}

%




\end{document}